\documentclass[a4paper,referee]{raa}     

\usepackage{graphicx,times}
\usepackage{natbib}
\usepackage{amssymb,amsmath}
\usepackage{threeparttable}
\usepackage{orcidlink}
\usepackage{upgreek}

\makeatletter

\newcommand{\Rmnum}[1]{\expandafter\@slowromancap\romannumeral #1@}
\makeatother

\modulolinenumbers[1] 
\linenumberdisplaymath 

\graphicspath{{./}{figures/}}
\newcommand{\facilities}[1]{\noindent\emph{Facilities:} #1}
\newcommand{\software}[1]{\noindent\emph{Software:} #1}

\begin{document}

\title{The Mini-SiTian Array: Evaluation Camera System}
\volnopage{ {\bf 20XX} Vol.\ {\bf X} No. {\bf XX}, 000--000}
\setcounter{page}{1}

\author{Yu Zhang\inst{1}\orcidlink{0000-0003-2536-2641}, Lin Du\inst{2,1}\orcidlink{0000-0002-4915-4137}, Yi Hu\inst{1}\orcidlink{0000-0003-3317-4771}, Yang Huang\inst{2,1}\orcidlink{0000-0003-3250-2876}, Ying Wu\inst{1,3}, Min He\inst{1}\orcidlink{0000-0001-6139-7660}, Hongrui Gu\inst{2,1}\orcidlink{0009-0007-5610-6495}, Haiyang Mu\inst{1}, Xunhao Chen\inst{2,1}, Hong Wu\inst{1,2}}


\institute{Key Laboratory of Optical Astronomy, National Astronomical Observatories, Chinese Academy of Sciences, Beijing 100101, People's Republic of China; {\it yzhang@bao.ac.cn}\\
\and School of Astronomy and Space Science, University of Chinese Academy of Sciences, Beijing 101408, People's Republic of China \\
\and Institute for Frontiers in Astronomy and Astrophysics, Beijing Normal University, Beijing 100875, China \\
}

\abstract{
The Mini-SiTian project, which is the pathfinder for the SiTian project, utilizes three 30 cm telescopes equipped with commercial CMOS cameras (ZWO ASI6200MM Pro) to simulate large-area time-domain survey. Due to the avoidance of the traditional mechanical shutter, the CMOS camera is favorable in time-domain survey projects. In the future, the SiTian telescope array will employ a two-by-two scientific-grade mosaic CMOS camera to survey a 10,000-degree square area every 30 minutes. Therefore, the performance of CMOS directly determines the detection capability of SiTian telescopes for transient sources, and a comprehensive understanding of the performance of CMOS cameras is crucial.
In this research, laboratory testing was conducted to thoroughly evaluate three cameras by assessing several critical parameters, including bias stability, dark current, pixel anomalies, linearity, gain, and read noise.
We find exceptional short-term bias stability with standard deviations below 0.02 ADU, negligible dark current of approximately 0.002 e$^{-}$ pixel$^{-1}$ s$^{-1}$ at $0^\circ\text{C}$, and excellent linearity with nonlinearity consistently below $\pm$ 0.5\%, and a small proportion (0.06\% to 0.08\%) of pixels with anomalous responses.
Furthermore, our analysis demonstrates uniform gain values across all cameras, ranging from 0.252 to 0.255 e$^{-}$ ADU$^{-1}$, with low readout noise, measured to be below 1.6 e$^{-}$ using conventional methods.
We also propose a novel method for pixel-level gain and read noise calculation for CMOS sensors, which revealed a narrow gain distribution and a low median read noise of 1.028 e$^-$ for one of the cameras.
The laboratory testing of the ZWO ASI6200MM Pro cameras indicates their potential to meet the requirements of time-domain surveys for the Mini-SiTian project.
}

\authorrunning{Yu Zhang et al.}            
\titlerunning{The Mini-SiTian Array: Evaluation Camera System}  
\maketitle

%
\section{Introduction} \label{introduction}
Complementary Metal-Oxide Semiconductor (CMOS) technology emerged in the 1960s, alongside Charge-Coupled Device (CCD) technology \citep{bigas_review_2006}. This technology is based on integrated circuits comprising both n-type and p-type metal-oxide Semiconductor Field-Effect Transistors (MOSFETs).
Conventional CMOS image sensor has limitations, such as low quantum efficiency, high dark current, small dynamic range, and limited potential full well \citep{bigas_review_2006}. However, recent advancements have addressed many of these issues. Modern CMOS sensors exhibit several notable advantages, including compact size, lightweight design, cost-effectiveness, and high readout speed. These benefits are largely attributed to the integration of individual amplifiers within each pixel. This architectural innovation allows each pixel to independently amplify the signal from its photodiode, resulting in a higher signal-to-noise ratio and enhanced processing speed \citep{PCO_scmos_white_paper, 2021RMxAC..53..190K, alarcon_scientific_2023}. 

Due to its numerous advantages, CMOS technology is extensively utilized in digital cameras, mobile phones, and monitoring equipment. 
However, in optical astronomical observation, CCD maintains dominance due to its superior sensitivity for faint objects and high precision measurements. These advantages, along with excellent performance in time-domain astronomy for detecting variable phenomena, make CCDs ideal for diverse and demanding astronomical applications (e.g., \citealt{1998AJ....116.3040G}; \citealt{2012SPIE.8446E..6RM}; \citealt{Wang_2017}).
Consequently, telescopes employing CMOS sensors as detectors remain relatively uncommon.

However, compared to CCDs, CMOS detectors present more unique opportunities for astronomical observations. They reduce costs and offer a higher degree of integration, facilitating the formation of large arrays and enhancing sensitivity for detecting faint objects. Additionally, CMOS detectors possess higher readout speeds and frame rates, enabling the capture of astronomical transients with rapid variations. These attributes make CMOS technology particularly advantageous for time-domain astronomy, where 
fast and precise measurements are essential.

In recent years, the field of time-domain astronomy in the optical band has seen rapid development. As technology advances, higher demands are placed on observational capabilities. In parallel, CMOS detector technology has evolved significantly, and has been increasingly adopted in various astronomical applications, including both wide-field surveys and individual telescope systems. Several projects have demonstrated the effectiveness of CMOS detectors in astronomical observations.
\cite{qiu_evaluation_2013} conducted comprehensive tests on a scientific CMOS camera (DC-152Q-FI by Andor Technology) to characterize its performance parameters.
Their results demonstrated the stable bias, gain, readout noise, and photometric precision of the cameras.
\cite{2020AcASn..61...37D} demonstrated the excellent linearity (better than 99\%) and uniformity (better than 3\%) of commercial-grade large-format CMOS chips with global shutters, highlighting the growing capabilities of CMOS technology in astronomical applications. 
\cite{zhang_tsinghua_2020} developed the Time-domain Multi-band Transient Survey (TMTS) system, employed the QHY4040 CMOS detector. This system have successfully detected a variety of transient phenomena in their observations, showcasing the potential of CMOS technology in time-domain astronomy.
Further supporting the viability of CMOS detectors for precision astronomy, \cite{2024RAA....24e5009M} evaluated an observational system with a ZWO ASI6200MM Pro Camera and a 60cm telescope. Their study assessed the photometric and differential photometric precision of the system through various astronomical observations. 
The results were promising, with photometric precision reaching 0.02 mag and differential photometry precision of 0.004 mag. 
This work demonstrates that CMOS-based systems can achieve precision comparable to traditional CCD systems across optical wavelengths. 

These studies collectively demonstrate that CMOS detectors are not only suitable for wide-field surveys but are also being successfully integrated into various telescope systems for targeted observations. The adoption of CMOS technology in astronomy is driven by its advantages in high readout speed, low noise, and cost-effectiveness, making it an increasingly attractive option for both large-scale surveys and precision astronomical measurements. 

With the enhancement of telescope detection capabilities, an increasing number of transient sources are being discovered. 
The SiTian Project is one of next-generation time domain surveys, aiming at monitoring large sky areas  every 30 minutes with 60 one-meter-class telescopes to detect various types of optical transients \citep{2021AnABC..93..628L}, such as supernovae, flare variable, gamma-ray bursts, tidal disruption events, and optical counterparts of gravitational waves. 
The early detection of these transients provides valuable information for exploring their nature. However, the luminosity of some transients changes rapidly, making it crucial to use CMOS detectors to capture these early light variations.

The Mini-SiTian (MST) project (cite He et al.) serves as a pilot initiative for the SiTian project, accumulating valuable experience in operation, observation, and data processing for its successor. The Mini-SiTian project comprises three telescopes equipped with $g$, $r$, and $i$ filters, located at XingLong Observatory, National Astronomical Observatories, Chinese Academy of Sciences (NAOC). 
ZWO ASI6200MM Pro cameras\footnote{\url{https://astronomy-imaging-camera.com/product/asi6200mm-pro-mono}} developed by ZWO\footnote{\url{https://www.zwoastro.com}} are utilized as the detector for three telescopes. 
Its sensor is SONY IMX455 CMOS which is a back-illuminated CMOS image sensor with a resolution of 9576$\times$6388, and pixel size of 3.76$\mu$m, yielding a field of view of $2^\circ.29 \times 1^\circ.53$ (pixel scale $\sim 0^{\prime \prime}.86$). 
Detailed technical specifications of the camera \footnote{\url{https://i.zwoastro.com/zwo-website/manuals/ASI6200_Manual_EN_v1.4.pdf}} is shown in Table \ref{tab:6200 tech}.

In this study, we performed detailed laboratory evaluations of three cameras by exaMining several critical parameters, including bias stability, dark current, pixel anomalies, linearity, gain, and read noise.
Laboratory testing of the ZWO ASI6200MM Pro cameras plays a crucial role in evaluating the performance of CMOS cameras, serving as the initial step to ensure device reliability and optimize astronomical observations. These tests provide essential baseline information on the fundamental characteristics of the equipment, laying the groundwork for subsequent on-sky testing.
We specifically reference the on-sky tests conducted by Xiao et al. (2024; also submitted to this special issue), which demonstrate that the same three ZWO ASI6200MM Pro CMOS camera systems, mounted on the Mini-SiTian telescopes, achieve high-precision photometric performance under actual astronomical observing conditions. These field results further validate the performance characteristics of the camera observed in our laboratory tests.
The study of Xiao et al. (2024; also submitted to this special issue) highlights the potential of these CMOS detectors for wide-field optical time-domain surveys. It shows that the detectors can achieve photometric accuracy comparable to CCDs, with impressive astrometric precision of approximately 70-80 mas and photometric accuracy of about 4 mmag for bright stars. The MST project underscores the feasibility of using CMOS detectors for large-scale time-domain surveys, offering a cost-effective alternative to CCDs.

\begin{table}[!ht]
\centering
\caption{ZWO ASI6200MM Pro technical specifications} \label{tab:6200 tech}
\begin{threeparttable}
\begin{tabular}{cc}
\hline
Feature & Specifications \\
\hline
Sensor          & SONY IMX455 CMOS \\
Diagonal        & 43.3mm           \\
Resolution      & 9576$\times$6388 \\
Pixel Size      & 3.76$\mu$m           \\
Image area      & 36mm$\times$24mm \\
Shutter         & Rolling shutter  \\
Exposure Range  & 32$\mu$s-2000s       \\
Read Noise      & 1.5-3.5e$^-$         \\
QE peak         & 91$\%$           \\
Full well       & 51.4 ke$^-$           \\
ADC             & 16 bit            \\
$\Delta$T       & $35^\circ\mathrm{C}$ (based on $30^\circ\mathrm{C}$ ambient temperature) \\
\hline
\end{tabular}
\end{threeparttable}
\end{table}

\section{Methods} \label{sec:obs and data reduc}
To evaluate the performance of the ZWO ASI6200MM Pro camera and verify its technical specifications provided by manufacturers, we conducted a series of laboratory tests on bias stability, dark current, abnormal response pixels, linearity, gain, and read noise for three cameras equipped with Mini-SiTian telescopes. 

The laboratory is designed as a dark room, equipped with an experimental platform for various experimental equipments, including integrating spheres and light sources. 
The room is maintained in complete darkness to facilitate the acquisition of dark images without interference from external light sources. 
This ensures optimal conditions for testing the performance of imaging sensors under true dark-field conditions.
The light source is installed inside the integrating sphere, and the light is diffusely reflected by the interior walls of the sphere. This results in a uniform and stable illumination emerging from a circular aperture on the sphere. The light exiting near the circular aperture can be considered to exhibit uniform intensity and stability. 
The camera was positioned outside the integrating sphere, on the central axis of the circular aperture, approximately 10 cm away from the aperture.
In our tests, the requirement was for a stable light source, the uniformity was not a critical factor. Consequently, the precise positioning of the camera relative to the integrating sphere was not of vital importance. 

The ZWO ASI6200MM Pro camera offers a range of configurable acquisition parameters. Unless otherwise stated, the following parameters were utilized for all three cameras during the experimental procedure: the gain was set to 100, which corresponds to a gain value of around 0.25 e$^-$ ADU$^{-1}$ as indicated by the gain value in the FITS file header. The offset and USB bandwidth parameters were set to 50 and 40, respectively. The sensor cooling temperature was set to $0^\circ\mathrm{C}$. 

A series of images were taken in the sequence of bias, flat and dark frames after the temperature was stable. The detailed exposure plan is outlined in Table \ref{tab:exp plan}. 
For bias images, the light source was turned off, and the camera was covered with black cloth to ensure a completely dark environment. One hundred bias images were captured for each of the three cameras. 
For flat image acquisition, the CMOS sensor was illuminated by a stable light source emanating from the aperture of the integrating sphere. The brightness of the light source was adjusted such that the average counts of the central region of the flat image for 2s exposure fell within the range of 20,000 to 40,000 ADU. 
Flat images were captured with varying exposure times, with several exposure time intervals (as detailed in the Exposure Time Interval column of Table \ref{tab:exp plan}). Within each interval, different exposure times were selected based on a specific step length (as detailed in the Exposure Time Step column of Table \ref{tab:exp plan}), and a flat image was exposed for each selected exposure time.
The conditions for dark image acquisition were identical to those for bias images, and the exposure times of bias images were consistent with those for flat images. 

\begin{table}[!ht]
\centering
\caption{Exposure plan} \label{tab:exp plan}
\begin{threeparttable}
\begin{tabular}{cccc}
\hline
Type & Exposure Time Interval & Exposure Time Step & Environment \\
& (s) & (s) &  \\
\hline
Bias & 0                & --               & Dark environment              \\
Flat & 0.01-0.10        & 0.01             & Stable light \\
Flat & 0.1-0.9         & 0.1              & Stable light \\
Flat & 0.9-0.99         & 0.01             & Stable light \\
Flat & 0.99-0.999       & 0.001            & Stable light \\
Flat & 1                & --               & Stable light \\
Flat & 1.001-1.010      & 0.001            & Stable light \\
Flat & 1.01-4.00        & 0.01             & Stable light \\
Flat & 4.0-8.1          & 0.1              & Stable light \\
Dark & 0.01-0.1         & 0.01             & Dark environment              \\
Dark & 0.1-1            & 0.1              & Dark environment              \\
Dark & 1                & --               & Dark environment              \\
Dark & 1.001            & --               & Dark environment              \\
Dark & 1.002            & --               & Dark environment              \\
Dark & 2                & --               & Dark environment              \\
Dark & 5                & --               & Dark environment              \\
Dark & 10               & --               & Dark environment              \\
Dark & 20               & --               & Dark environment              \\
Dark & 50-600           & 50               & Dark environment              \\
\hline
\end{tabular}
\end{threeparttable}
\end{table}

\section{Results} \label{sec:results}
\subsection{Bias Stability Analysis and Frame Characteristics} \label{sec:bias stab}

\begin{figure} 
\includegraphics[angle=0,width=110mm]{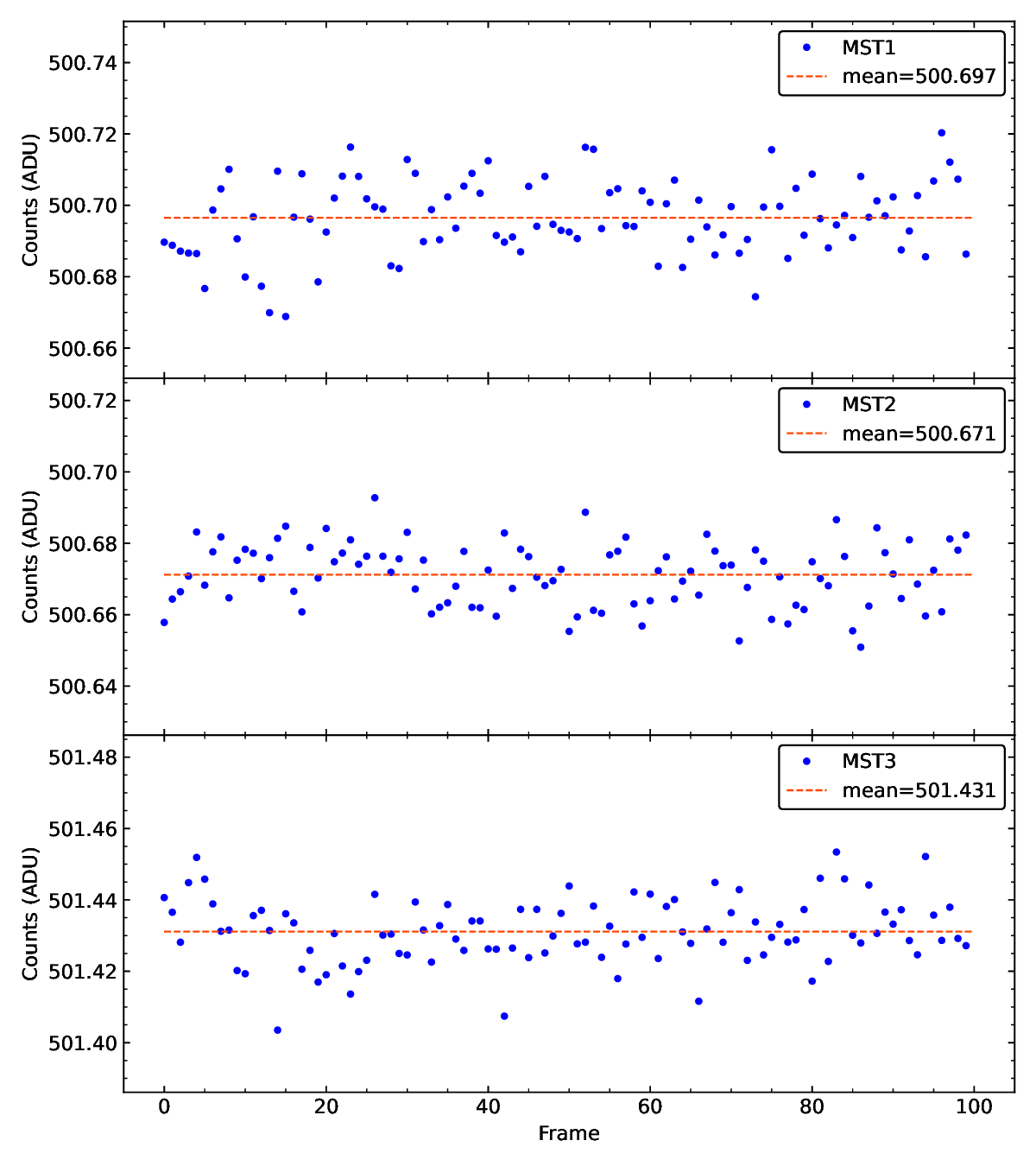}
\centering
\caption{Stability of the bias signal of three ZWO ASI6200MM Pro CMOS cameras of the Mini-SiTian array. Each blue dot represents the mean count within a 2000$\times$2000 pixels area at the center of the image for each bias image. The top-right corner marks values for mean, while the red dotted line indicates the overall mean of these values. }\label{fig:bias stab}
\end{figure}

Bias, also known as the bias frame, represents the inherent electronic noise and offset in the CMOS detector, captured with zero exposure time and closed shutter. It reflects the base level output of the sensor in the absence of any light or dark current. (not accuracy for the definition of bias frame, refer to handbook of ccd astronomy). 
In astronomical observations, all images require subtraction of the bias to isolate the photon signal and eliminate this systematic electronic offset. 
The stability of the bias is crucial, as it directly impacts the quality and accuracy of subsequent image processing steps, such as dark current subtraction and flat-field correction. 
In the context of this paper, the bias stability refers to the consistency of the bias signal over a time period under constant environmental conditions.
To evaluate the stability of the bias, we captured a 100 bias frames over a period of $\sim$ 20 minutes under controlled laboratory conditions. For each bias image captured by the three cameras, we calculated the mean of the counts within a 2000 $\times$ 2000 pixel area at the center of the image. This approach allows us to assess short-term variations between consecutive frames. A stable bias ensures a consistent noise floor, leading to more reliable and repeatable scientific measurements. In our analysis, we consider the bias to be stable as the variation in mean counts remains within $\pm$ 0.03 ADU of the overall average throughout the short observation period.
The results of our bias stability analysis for each camera are shown in Figure \ref{fig:bias stab}.
This figure illustrates the temporal evolution of the mean bias counts for each camera over $\sim$ 20 minutes.
The red dotted line in each panel represents the average of these 100 mean values, calculated using a sigma-clipping method. We employed a 3$\sigma$ clipping algorithm to mitigate the impact of potential outliers.
As for long-term stability of bias, \cite{2024RAA....24e5009M} found that the deviation mean counts of bias in one night was less than 0.08 ADU.
This results are crucial for understanding the consistency of the bias level across multiple acquisitions, which is fundamental for accurate image processing and scientific data reduction in astronomical observations.

\begin{figure} 
\includegraphics[angle=0,width=49mm]{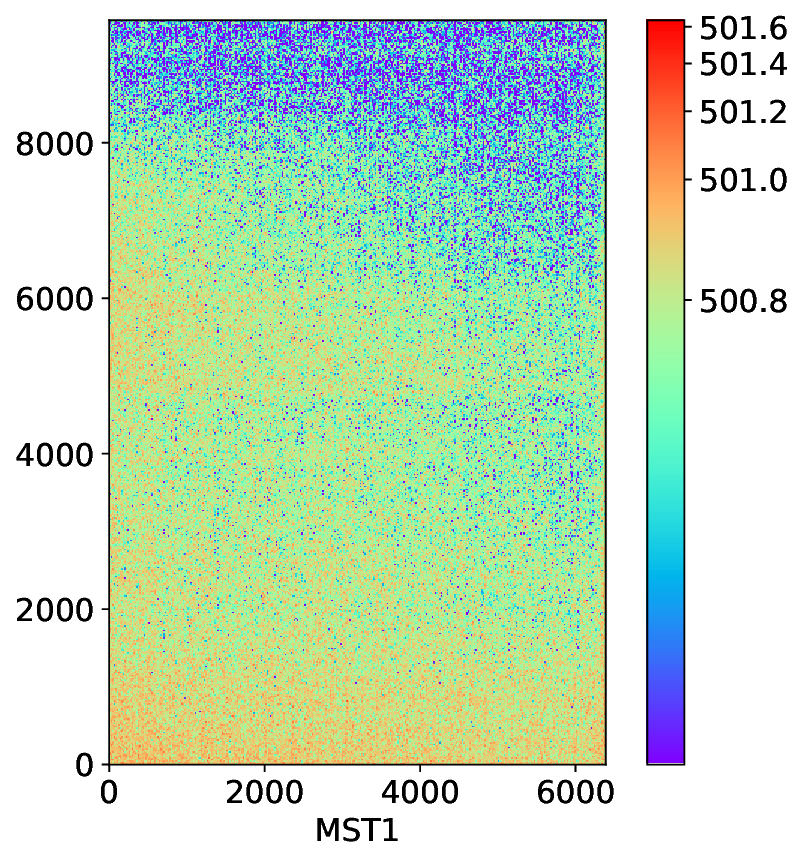}
\includegraphics[angle=0,width=49mm]{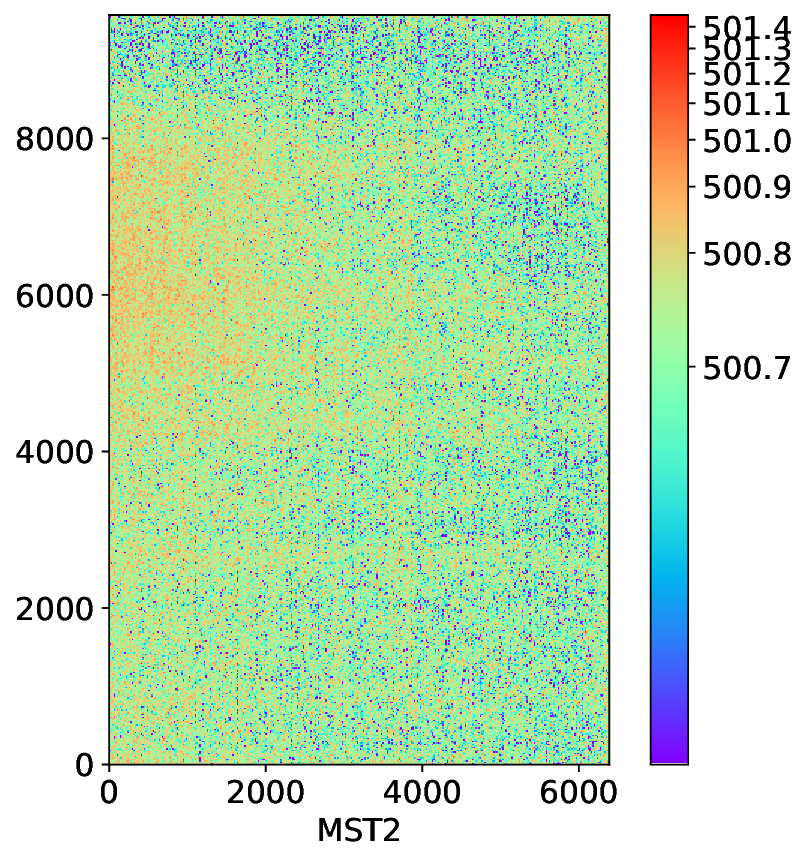}
\includegraphics[angle=0,width=49mm]{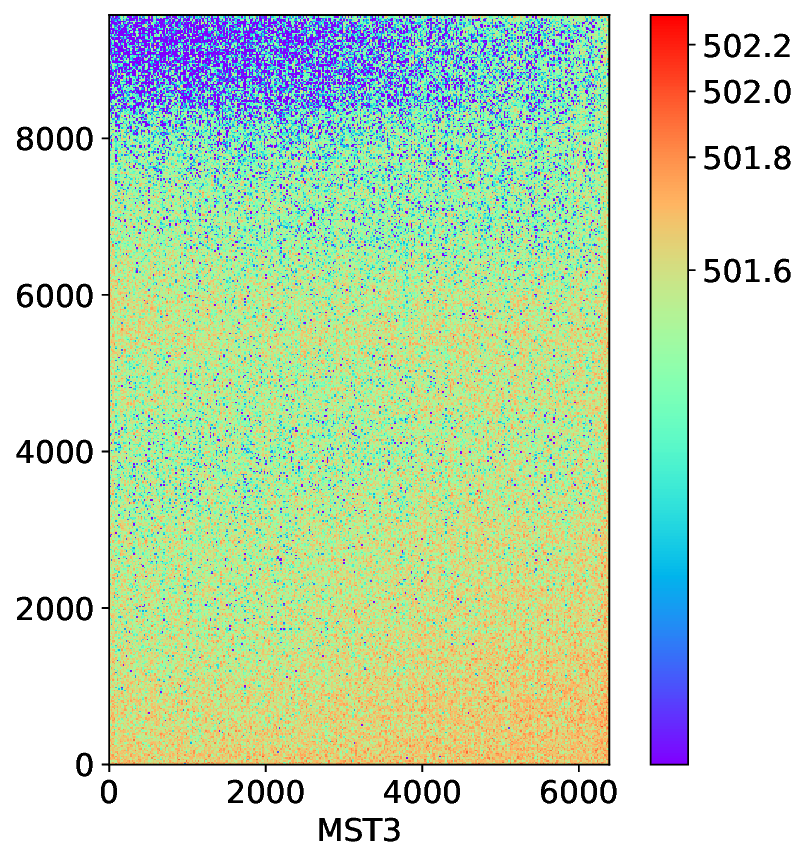}
\centering
\caption{Spatial distribution of pixel values in the combined bias frames, each derived from 100 individual bias images, for three ZWO ASI6200MM Pro CMOS cameras of the Mini-SiTian telescopes. The distribution is represented using a color-coded scheme, with the corresponding count values indicated by the color bar adjacent to each panel. Each panel represents the combined bias frame for one of the three cameras, which can be identified by the label on the x-axis of each panel. } \label{fig:bias}
\end{figure}

To investigate the characteristics of bias frames in depth, we performed mean stacking of 100 individual bias frames for each ZWO ASI6200MM Pro CMOS camera in our Mini-SiTian telescope array. The abbreviations MST1, MST2, and MST3 refer to telescope 1, 2, and 3 of the Mini-SiTian array, respectively. Figure \ref{fig:bias} presents the resulting combined bias frames for these three cameras.
Analysis of these combined frames reveals several notable patterns, which can be largely attributed to the Fixed Pattern Noise (FPN; \citealt{ElGamal1998ModelingAE}) characteristic of CMOS sensors. FPN is a type of noise that remains constant across different exposures and is especially prominent in CMOS sensors due to their architecture. In the context of bias frames, we are primarily observing the effects of Dark Signal Non-Uniformity (DSNU; \citealt{jimenez2012high}), a component of FPN. It's important to note that while these patterns are consistent and detectable, the numerical differences they represent are quite small:
(1) Gradient Pattern: All three cameras exhibit a subtle gradient in their bias levels. In MST1 and MST3, the count values generally decrease from the bottom to the top of the frame, while MST2 shows an even less pronounced gradient.
(2) Corner Effects: In MST1 and MST3, the two corners along the bottom edge of the frames display slightly higher count values compared to other areas. This effect is less noticeable in MST2. Such corner effects are another aspect of FPN. 
(3) Horizontal Structures: All three cameras show faint horizontal structures parallel to the short edge of the frame. These structures appear as subtle horizontal striations or banding patterns. This is a common form of FPN in CMOS sensors, often associated with the row-by-row readout process. 
The presence of these subtle gradients, corner effects, and horizontal structures, all manifestations of FPN, underscores the importance of proper bias correction in the data reduction process. 

Furthermore, the distribution of mean count values for each pixel in these bias frames deviates from a Gaussian (normal) distribution, as shown in Figure \ref{fig:bad pix1}. This non-Gaussian nature can be attributed to the DSNU component of FPN. The distributions for all three cameras exhibit asymmetry, with a long tail extending towards higher ADU values, and appear broader than a standard Gaussian distribution. 
These characteristics likely stem from the unique architecture of CMOS sensors, particularly their independent pixel readout system. In CMOS sensors, each pixel is equipped with its own readout circuitry. This design offers advantages in terms of speed and flexibility. However, it can also exacerbate the effects of DSNU. 
The independent readout structure leads to inherent variations in response and noise characteristics across the sensor. These variations significantly contribute to the observed non-Gaussian distribution of pixel values in the bias frames. 
Furthermore, the specific readout mechanisms and electronics of the ZWO ASI6200MM Pro camera modulate these characteristics, likely influencing both the asymmetry in pixel value distributions and the distinct patterns observed in combined bias frames. 

\subsection{Anomalous Response Pixels} \label{sec:abnpix}
In CMOS sensors, there exist pixels with abnormal responses, which can manifest as significant deviations in the pixel count values, appearing either excessively high or low compared to their surroundings, or exhibiting unusually large standard deviations. 
These anomalies can be attributed to various factors, including manufacturing defects, structural inconsistencies within the pixel, variations in dark current, non-ideal behavior of the readout circuitry, or degradation over time. 
The `Salt and Pepper' effect, known as Random Telegraph Noise (RTN), is a common manifestation of this issue. It refers to pixels that exhibit either very high (salt) or very low (pepper) values, which can appear as random speckles in the image. 
This peculiar phenomenon has been found in QHY411M \citep{alarcon_scientific_2023}, and is evident on some pixels, where the distribution of their counts exhibit either a bimodal or trimodal pattern. 
The presence of such anomalous response pixels could potentially affect photometric accuracy to the extent that they require precise identification and annotation.

\begin{figure} 
\includegraphics[angle=0,width=150mm]{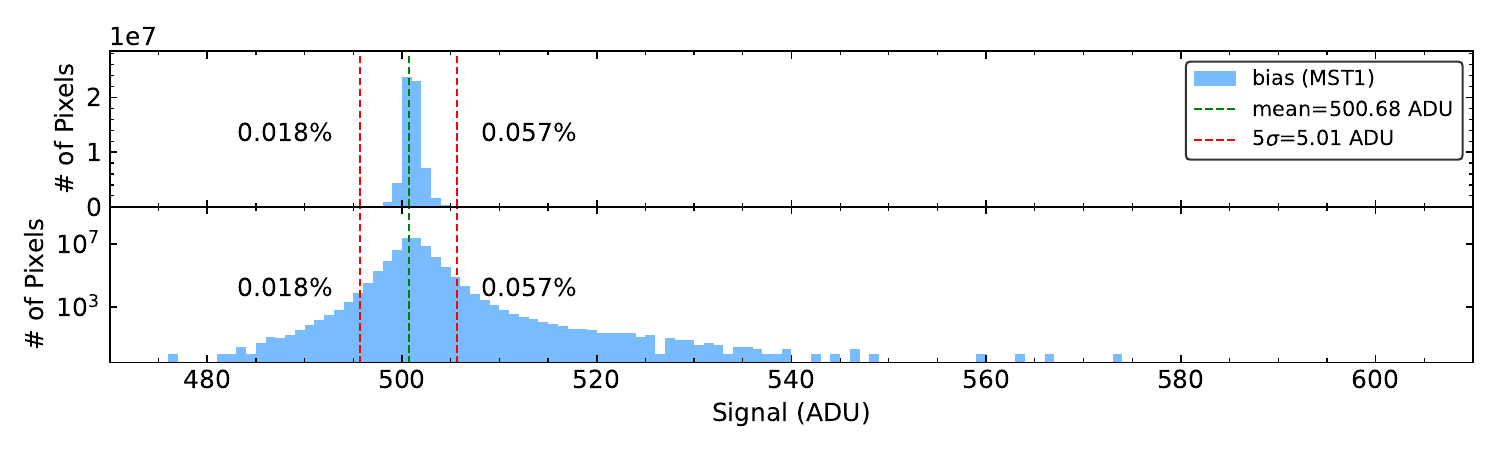}
\includegraphics[angle=0,width=150mm]{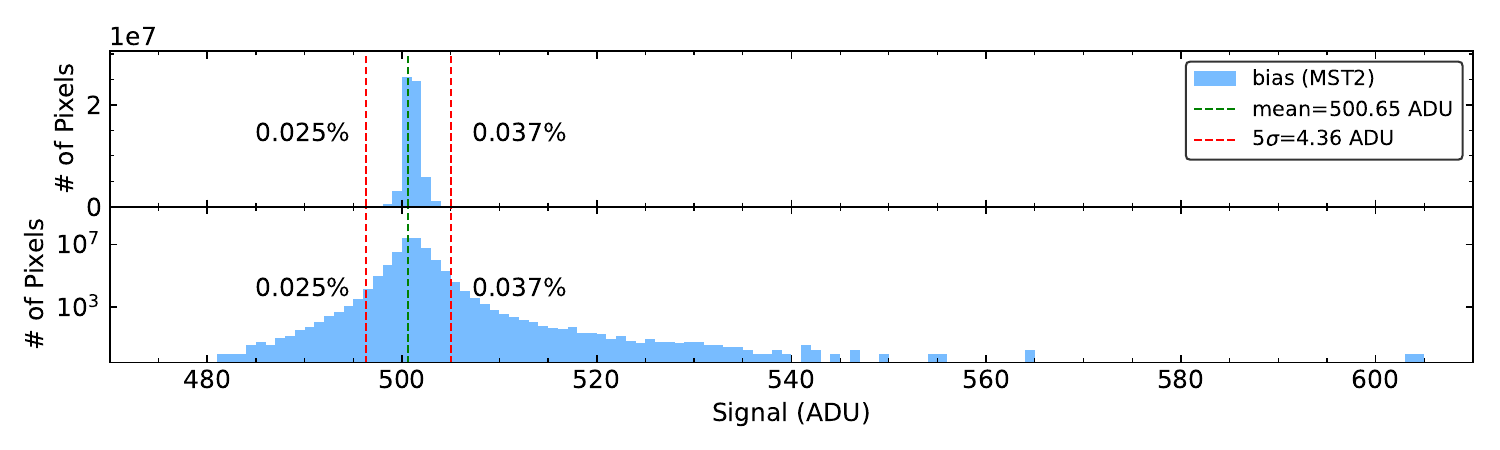}
\includegraphics[angle=0,width=150mm]{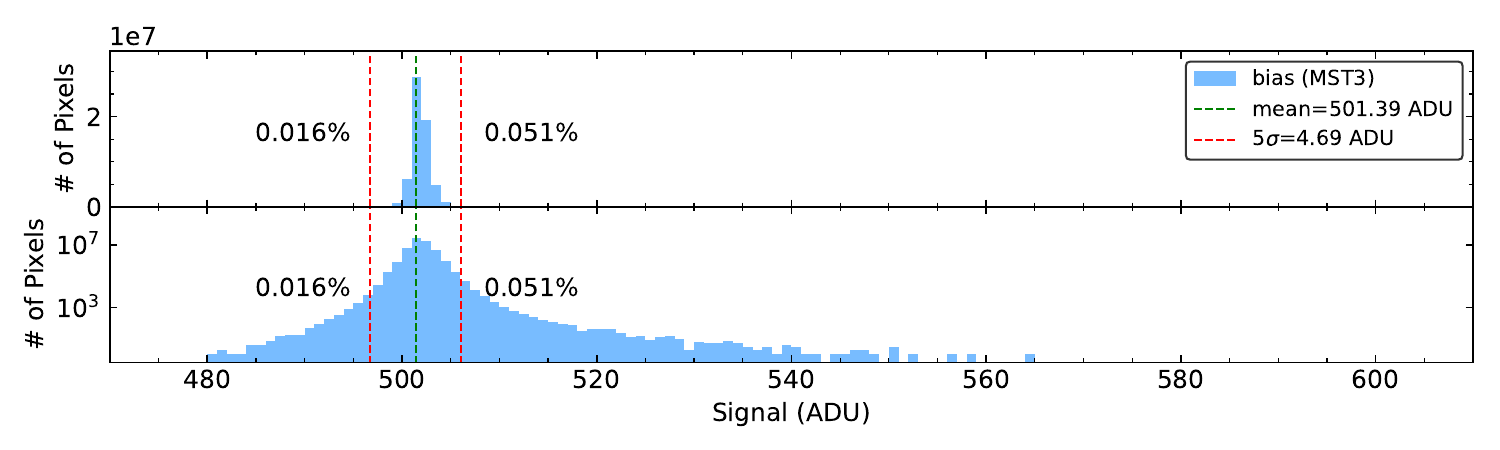}
\centering
\caption{Distribution of mean count values for each pixel of three ZWO ASI6200MM Pro CMOS cameras attached to the three Mini-SiTian telescopes, derived from 100 bias images. The dashed green lines indicate the mean value, and the red dashed lines marks the 5$\sigma$ thresholds. The top section of each panel uses a linear vertical axis, while the bottom section employs a logarithmic scale to highlight the tails of the distribution. The legend provides values for mean and 5$\sigma$. } \label{fig:bad pix1}
\end{figure}

To assess the prevalence of anomalous pixels, we analyzed a series of 100 bias images to calculate the mean and standard deviation of the count for each pixel.
The resulting distributions of the mean count of each pixel for the three cameras are presented in Figure \ref{fig:bad pix1}. 
Each panel represents a distinct camera. 
The percentages of pixels exhibiting extreme responses, falling beyond the 5$\sigma$ limit, are highlighted within the figure. 
The top panel corresponds to the camera attached to MST1, with approximately 0.018\% of pixels having mean counts below mean$-5\sigma$ and another 0.057\% exceeding mean$+5\sigma$. 
Similarly, the middle panel shows the results for the camera of MST2, where about 0.025\% of pixels fall below mean$-5\sigma$, while 0.037\% surpass mean$+5\sigma$. 
Lastly, the bottom panel represents the camera of MST3, with 0.016\% of pixels exhibiting counts below mean$-5\sigma$ and 0.051\% above mean$+5\sigma$. 
These findings indicate that the proportion of pixels displaying unusually high or low responses ranges from 0.06\% to 0.08\%, demonstrating the presence of noise sources such as RTN. 

\begin{figure} 
\includegraphics[angle=0,width=150mm]{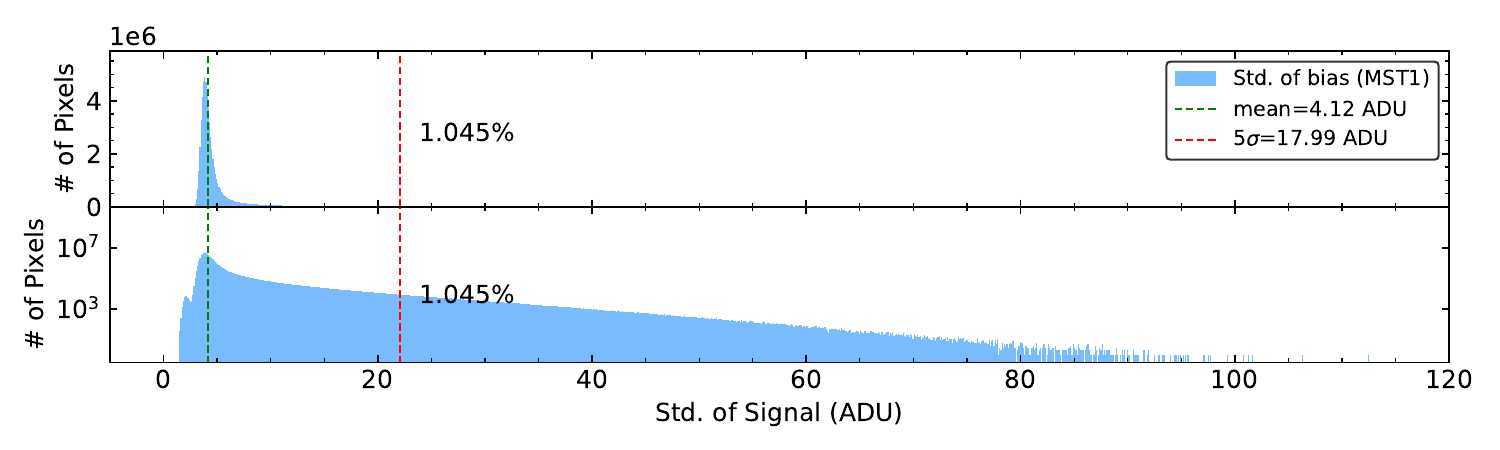}
\includegraphics[angle=0,width=150mm]{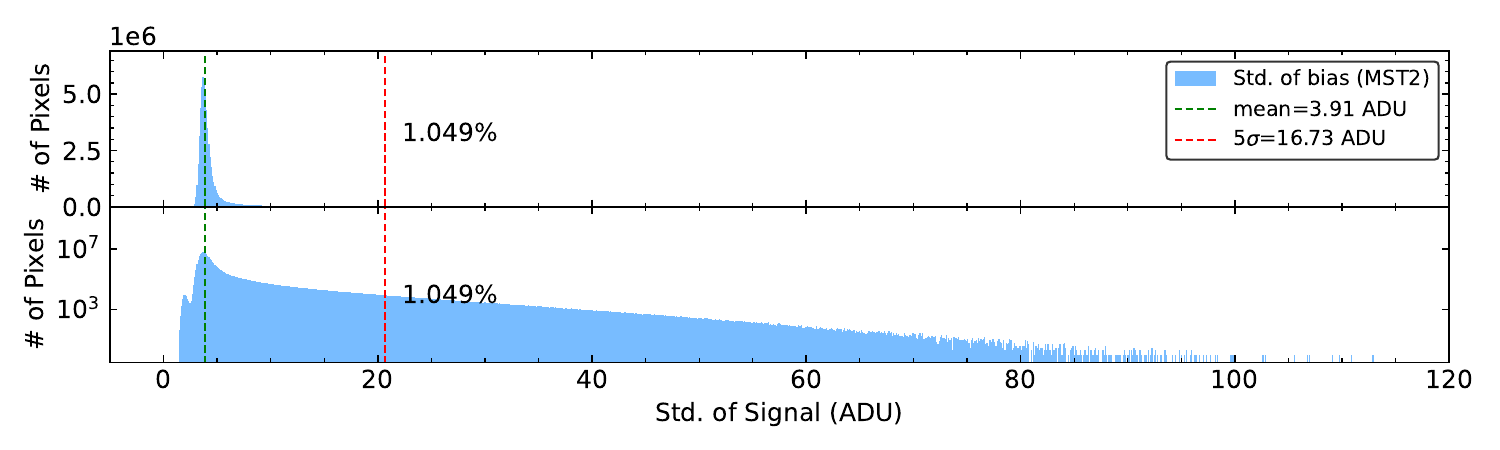}
\includegraphics[angle=0,width=150mm]{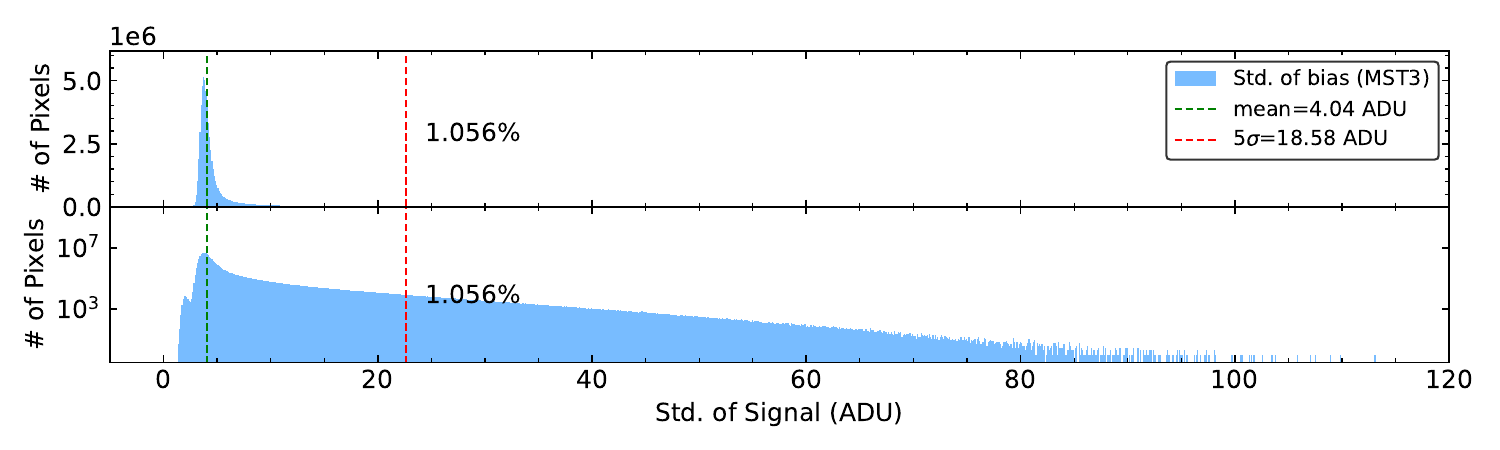}
\centering
\caption{Distribution of standard deviation values for each pixel in ZWO ASI6200MM Pro CMOS cameras attached to the three Mini-SiTian telescopes. The dashed green line indicates the mean standard deviation, and the red dashed line marks the 5$\sigma$ threshold. The top section of each panel uses a linear vertical axis, while the bottom section employs a logarithmic scale to highlight the tails of the distribution. The legend provides values for mean and 5$\sigma$. } \label{fig:bad pix2}
\end{figure}

Additionally, The standard deviation of each pixel was calculated to identify those with significant variations in response. These results are displayed in Figure \ref{fig:bad pix2}, which consists of three panels representing the three cameras. 
As seen in the figure, the mean value and the 5$\sigma$ threshold of these values are depicted by the dashed green and red lines, respectively. 
The percentage of pixels with standard deviations beyond the 5$\sigma$ threshold is approximately 1.05\% across all three cameras.
The top panel represents camera of MST1 with a mean standard deviation of 4.12 ADU and a 5$\sigma$ threshold at 22.11 ADU. The middle panel corresponds to camera of MST2 with a mean standard deviation of 3.91 ADU and a 5$\sigma$ threshold at 20.64 ADU. Finally, the bottom panel illustrates the results for camera of MST3, featuring a mean standard deviation of 4.04 ADU and a 5$\sigma$ threshold at 22.62 ADU.
It should be noted that none of the cameras exhibit pixels with a constant value, as there are no instances of zero standard deviation observed in any of the three cameras. 

\begin{figure} 
\includegraphics[angle=0,width=49mm]{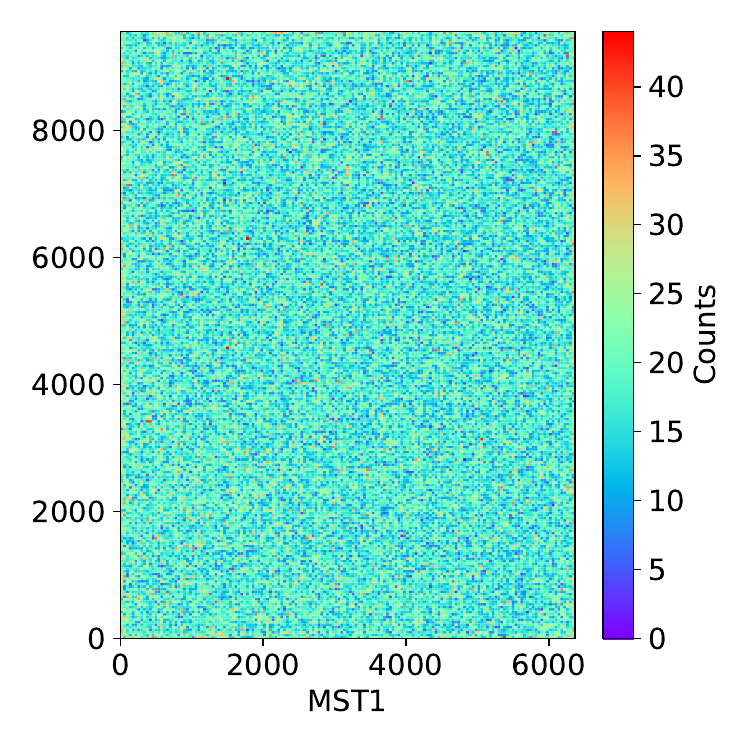}
\includegraphics[angle=0,width=49mm]{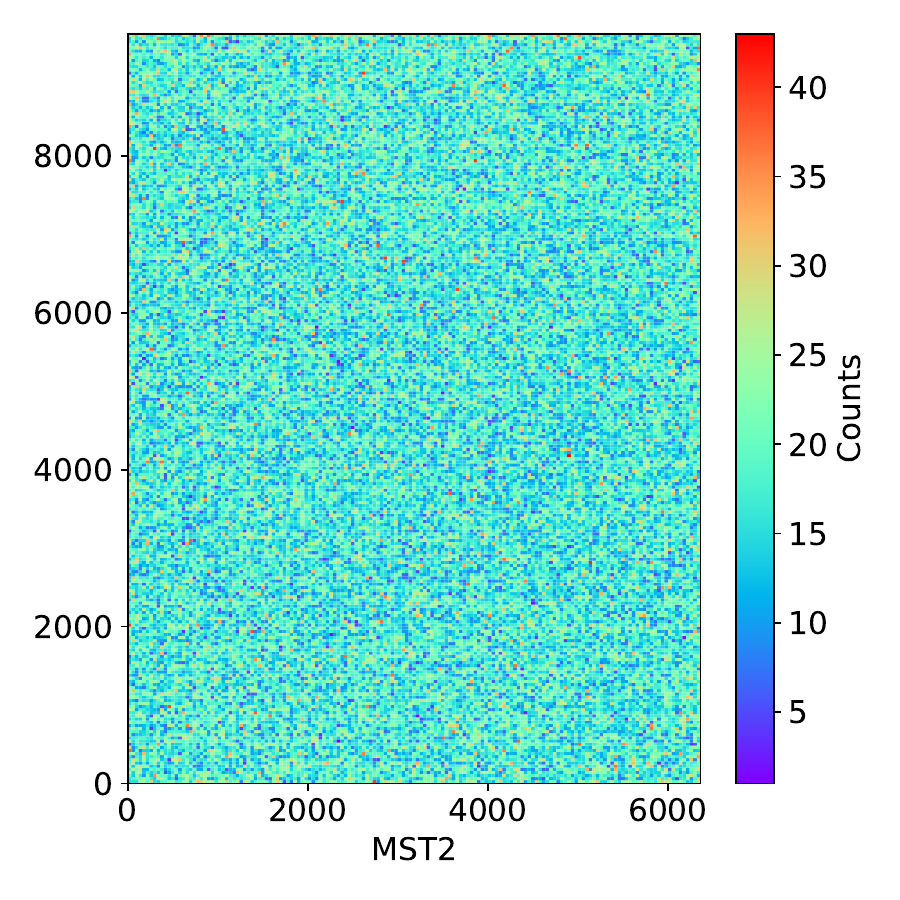}
\includegraphics[angle=0,width=49mm]{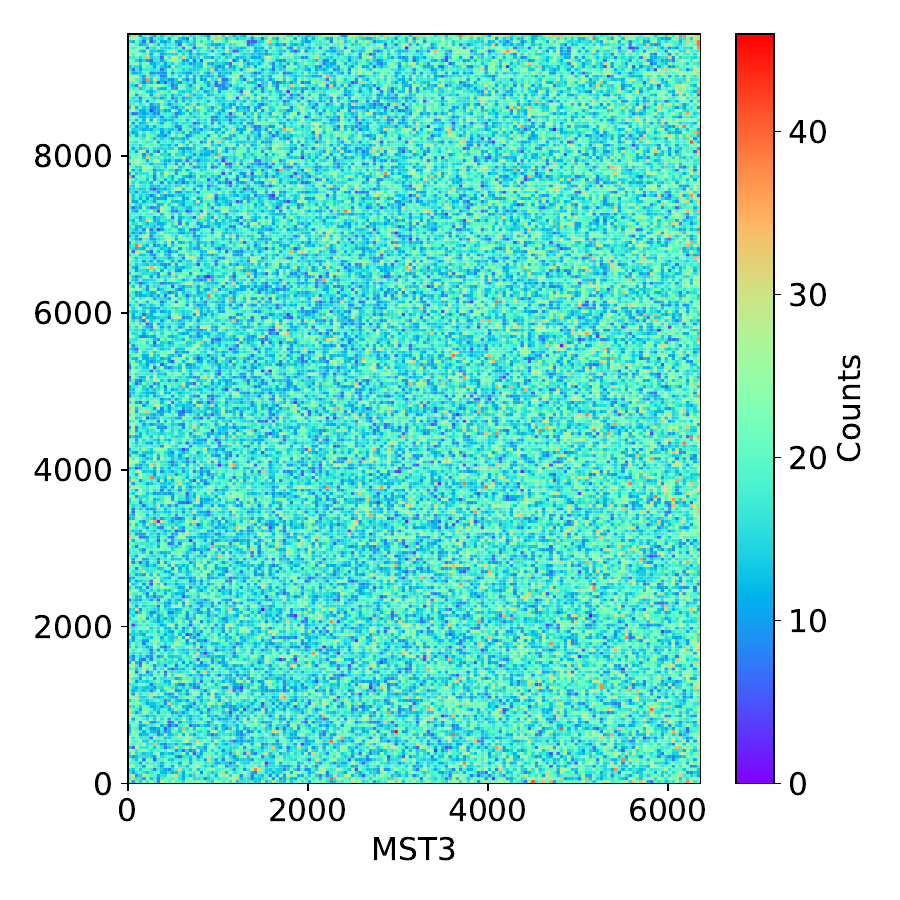}
\centering
\caption{Spatial distribution of anomalous pixel counts within 20 $\times$ 20 pixel subarea across the entire sensor for the three ZWO ASI6200MM Pro CMOS cameras of the Mini-SiTian telescope. The distribution is visualized using a color-coded scheme, with corresponding count values indicated by the color bar adjacent to each panel. Each panel represents the distribution for one of the three cameras, which can be identified by the label on the x-axis of each panel. } \label{fig:badpos}
\end{figure}

Considering the spatial distribution of the identified anomalous pixels, Figure \ref{fig:badpos} offers insights into their locations in the CMOS sensor. 
The three panels in Figure \ref{fig:badpos} illustrate the anomalous pixel distribution across the entire sensor area for each of the three cameras in the Mini-SiTian telescope system.
A color-coded representation enables a visual comparison of count differences among various regions. It shows that these anomalous response pixels tend to cluster around the edges and corners of the CMOS sensor, indicating a possible correlation between their abnormal response and their proximity to the sensor boundaries. This observation may suggest potential issues related to manufacturing defects or mechanical stress affecting the peripheral regions of the detector array.

\begin{figure} 
\includegraphics[angle=0,width=49mm]{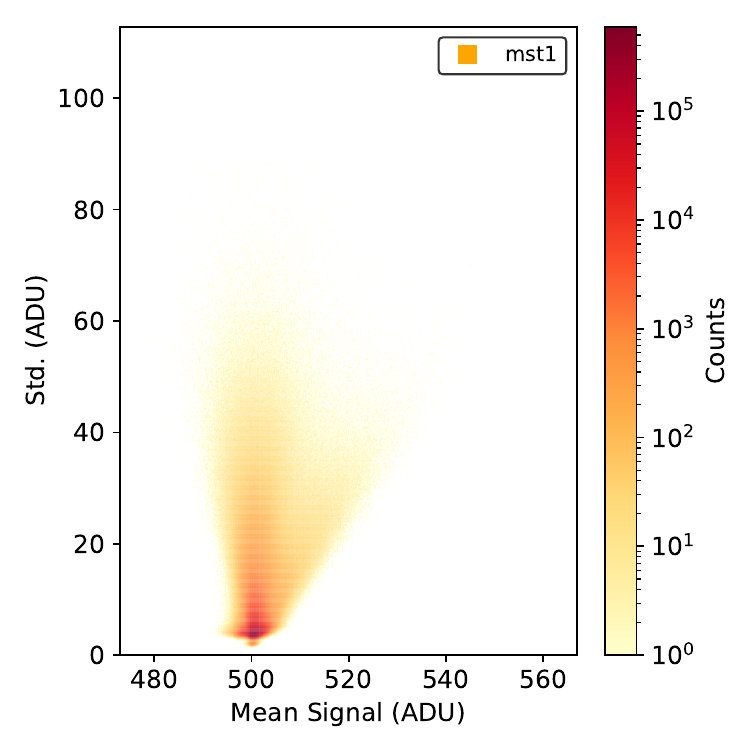}
\includegraphics[angle=0,width=49mm]{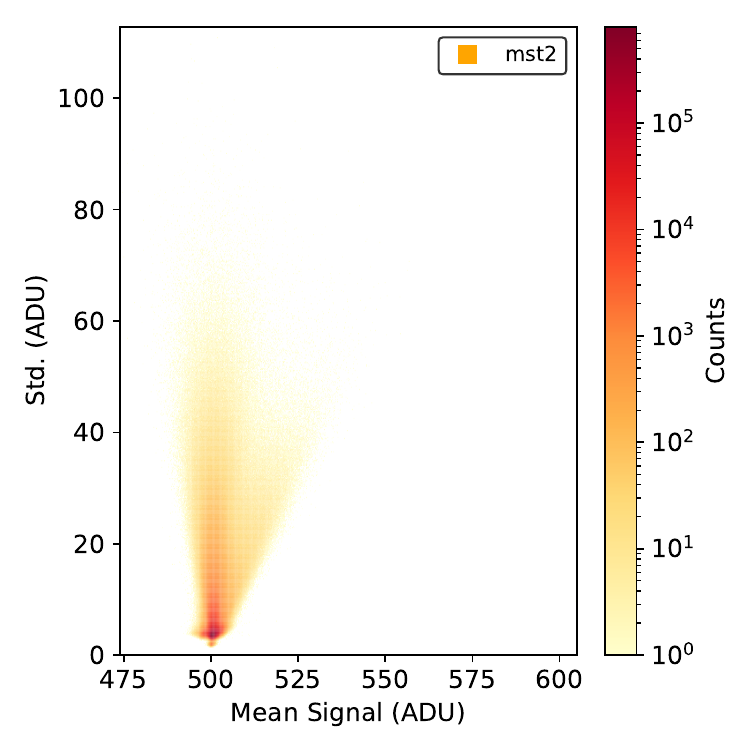}
\includegraphics[angle=0,width=49mm]{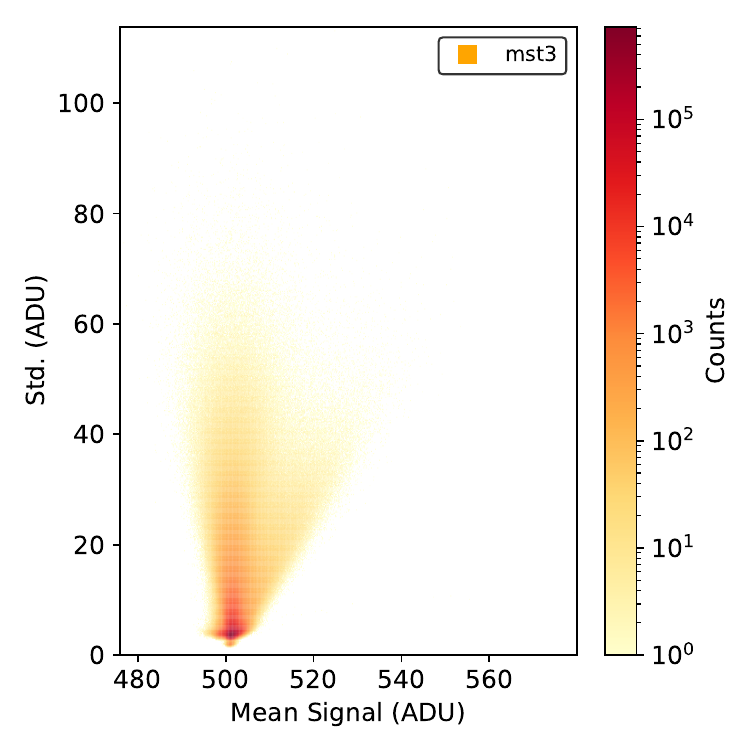}
\centering
\caption{Two-dimensional distribution of mean signal and standard deviation for each pixel of ZWO ASI6200MM Pro CMOS cameras of the Mini-SiTian telescope. } \label{fig:std}
\end{figure}

In Figure \ref{fig:std}, we present the two-dimensional distribution of the mean signal and standard deviation for each pixel, derived by combining the information from Figure \ref{fig:bad pix1} and Figure \ref{fig:bad pix2}. This visualization provides insight into the statistical properties of the pixel responses across the entire sensor area.

It becomes evident that the majority of pixels exhibit mean values ranging from 495 ADU to 505 ADU with standard deviation between 3 ADU and 5 ADU. 
This central region represents the typical behavior expected under normal conditions. 
However, two distinct branches of pixels deviate from this normal distribution, offering clues about potential anomalies in the performance of CMOS sensors. 
One branch is at the upper end of the normal pixel distribution, characterized by widespread dispersion in standard deviation. 
The other branch is on the upper right of the normal pixel distribution, with both the mean and standard deviation falling far outside the average range for most pixels.
Pixels distributed in both branches may exhibit a distinct `Salt and Pepper' effect. 

We have marked these anomalous response pixels to reduce their impact on metering accuracy during data processing. 
Future work should focus on exploring the factors contributing to these anomalies and devising methods to mitigate them effectively. 

\subsection{Dark Current} \label{sec:dark}
\begin{figure} 
\includegraphics[angle=0,width=110mm]{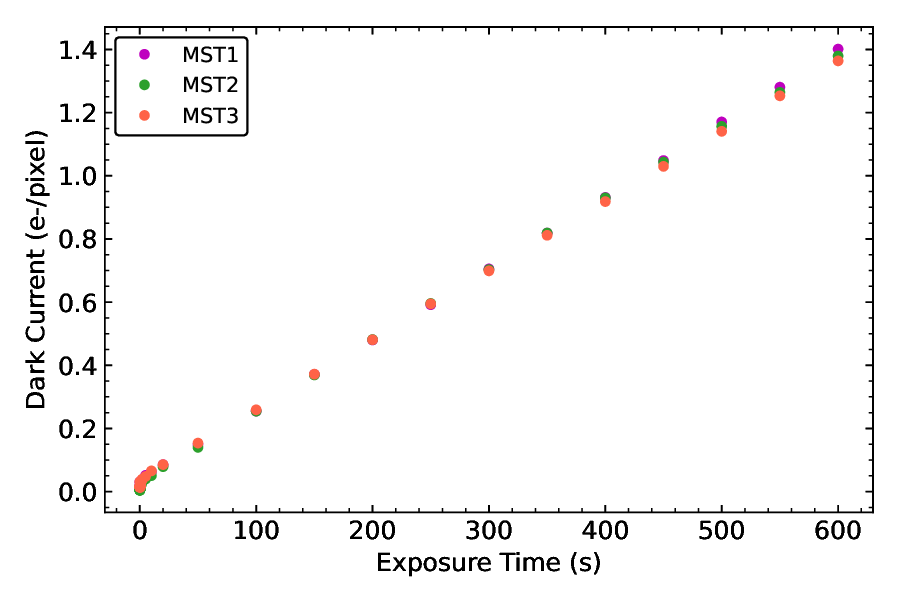}
\centering
\caption{Dark current versus exposure time of three ZWO ASI6200MM Pro CMOS cameras of Mini-SiTian telescopes. The orange, green, and red dots indicate the dark current (in units of electron per pixel) against exposure time (in units of seconds) for the ZWO ASI6200MM Pro CMOS cameras attached to the Mini-SiTian telescope1 (MST1), the Mini-SiTian telescope2 (MST2), and the Mini-SiTian telescope3 (MST3), respectively, showing a linear trend as exposure time increases, indicating a roughly proportional relationship between dark current and exposure duration. } \label{fig:dark current}
\end{figure}

Dark current primarily stems from thermal excitation. At ambient temperatures, electrons within semiconductor materials gain enough energy through thermal fluctuations to transition from the valence band to the conduction band, generating noise in astronomical imaging. 
To ensure high-quality observations, especially when capturing dark sources, it is crucial to have low dark current. 
A series of dark frames were collected over exposure times ranging from 0.01 seconds to 600 seconds, and subsequent subtraction of the master bias frames was performed to evaluate the dark current levels.
Analysis revealed a roughly linear relationship between dark current and exposure time (Figure \ref{fig:dark current}), with an average rate of approximately 0.002 e$^-$ pixel$^{-1}$ s$^{-1}$ at $0^\circ\text{C}$. 
While this rate is relatively low, the complex distribution of dark current across pixels requires careful consideration in data processing.

\begin{figure} 
\includegraphics[angle=0,width=140mm]{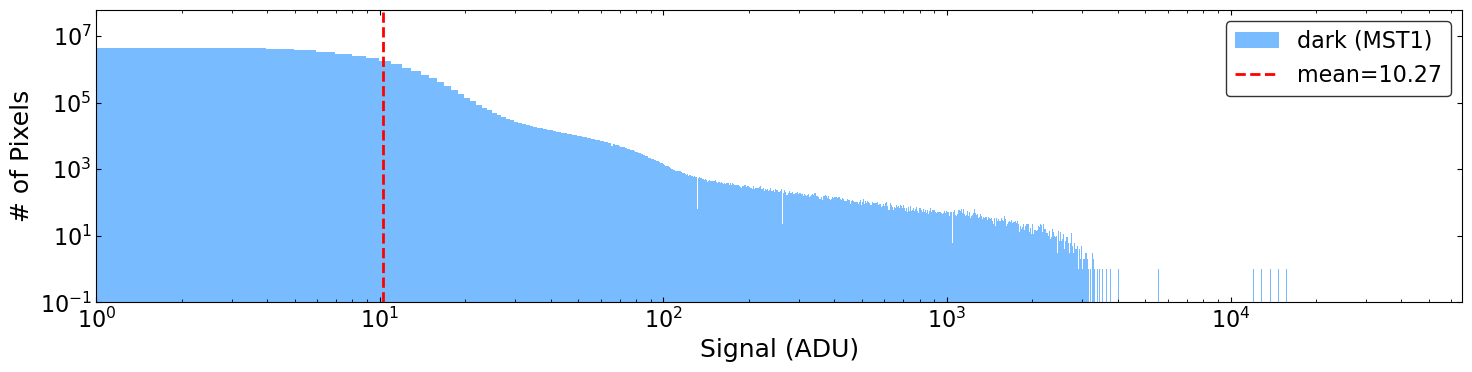}
\includegraphics[angle=0,width=140mm]{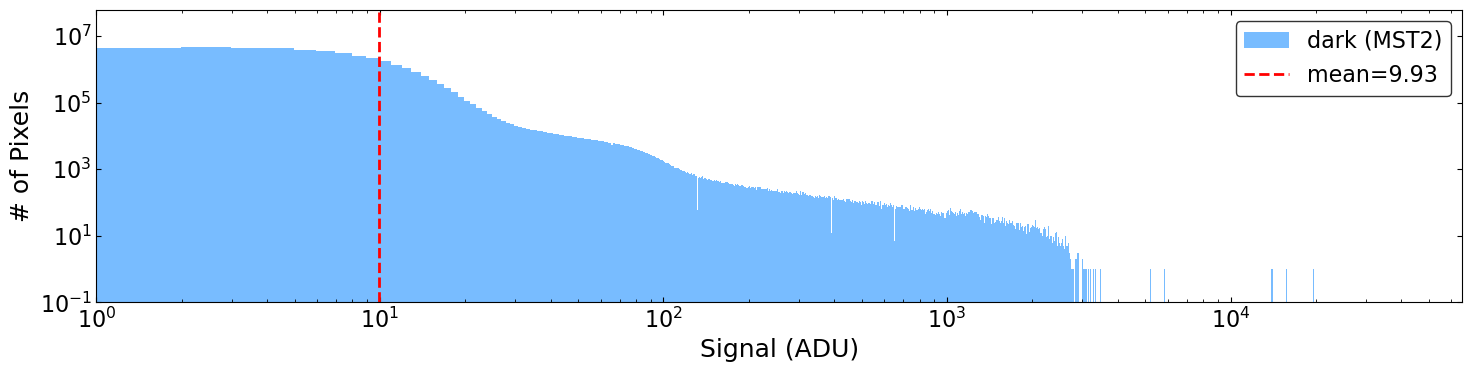}
\includegraphics[angle=0,width=140mm]{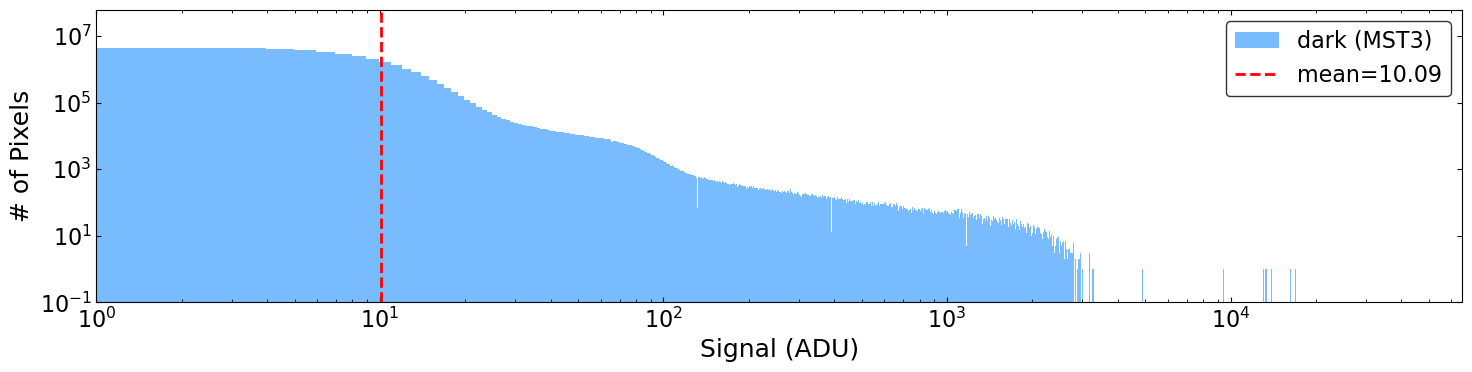}
\centering
\caption{Distribution of dark current values for the ZWO ASI6200MM Pro CMOS cameras in the Mini-SiTian Array Camera System. The histograms show the number of pixels (on a logarithmic scale) versus the dark current signal (in ADU) for MST1, MST2, and MST3. The red dashed lines indicate the mean dark current values for each camera. Multi-modal distribution with a primary peak near the mean and several secondary peaks at higher ADU values. } \label{fig:dark_distribution}
\end{figure}

Further investigation into the dark current distribution in our ZWO ASI6200MM Pro CMOS cameras revealed an interesting pattern that deviates from the commonly observed log-normal distribution in some CMOS sensors (e.g., \citealt{Konnik2014HighlevelNS}; Figure \ref{fig:dark_distribution}). 
The histograms show the distribution of pixel count values in 600-second dark frames after bias subtraction for all three cameras of MST1, MST2, and MST3. This distribution exhibits several distinct features:
(1) A primary peak: the distribution shows a prominent peak below the mean value, representing the typical dark current level for most pixels.
(2) Long tail: there is a tail extending towards higher ADU values, indicating the presence of a small number of pixels with higher dark current.
(3) Secondary peaks: the distribution suggests the potential existence of distinct populations of pixels with elevated dark current levels. 
This distribution pattern may be attributed to the specific design of the CMOS sensor used in these cameras and the underlying FPN. 

Understanding this distinctive dark current distribution is crucial for developing accurate noise reduction algorithms and optimizing the performance of the ZWO ASI6200MM Pro cameras in the Mini-SiTian Array Camera System. The cameras demonstrate good performance with a low average dark current rate of approximately 0.002 e$^-$ pixel$^{-1}$ s$^{-1}$ at $0^\circ\text{C}$, which enhances their ability to detect faint celestial objects and improves image quality, particularly during long exposures. However, the complex distribution indicates that some pixels may contribute more significantly to noise than others. This variability needs to be accounted for in data reduction and analysis processes to ensure optimal image quality and scientific accuracy.
Future work will focus on characterizing this distribution in detail, potentially modeling it using a mixture of probability distributions or investigating the physical origins of the observed pattern. Additionally, developing advanced, pixel-specific correction methods may further improve the overall performance of the camera system.

\subsection{Linearity}
\begin{figure} 
\includegraphics[angle=0,width=74mm]{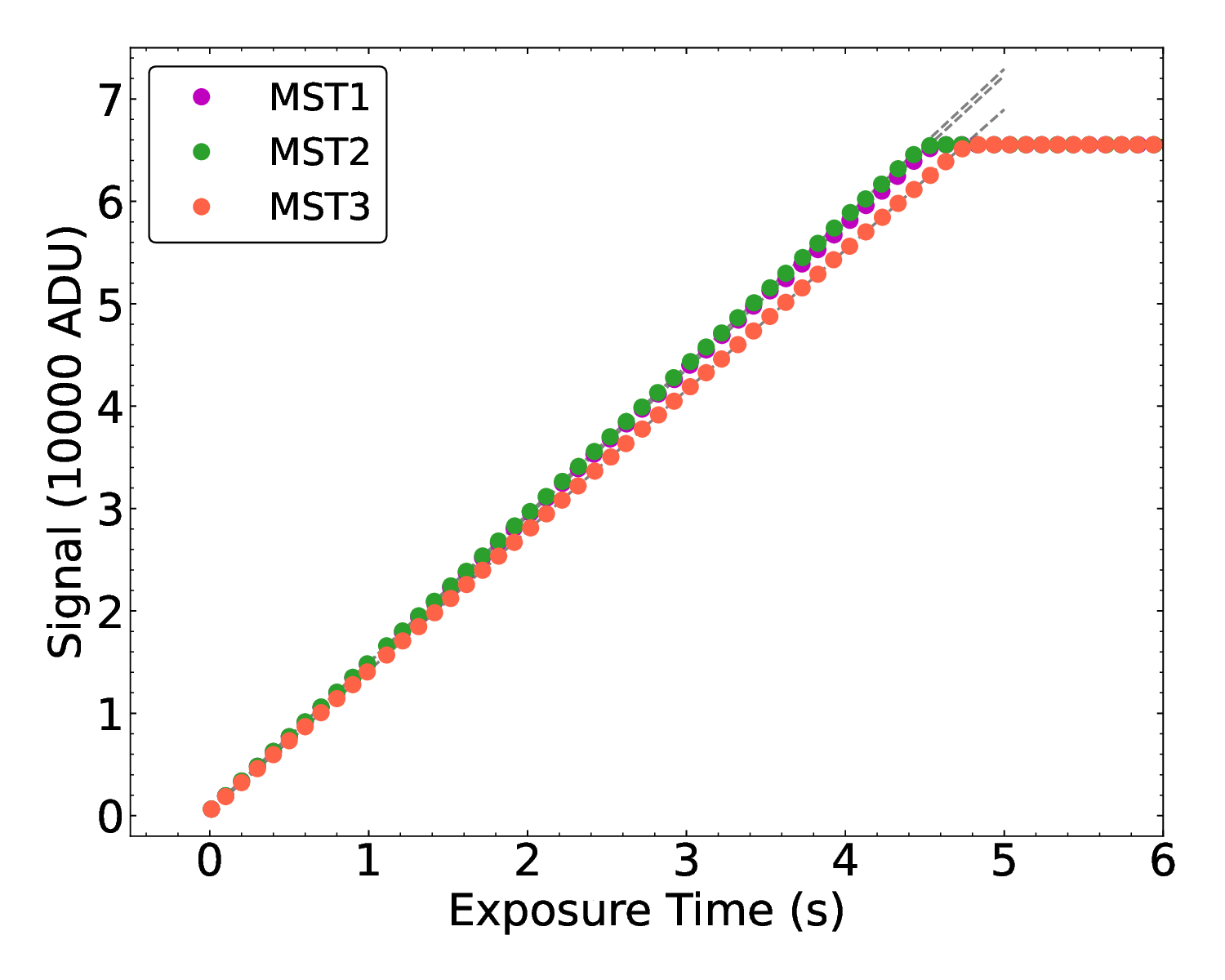}
\includegraphics[angle=0,width=74mm]{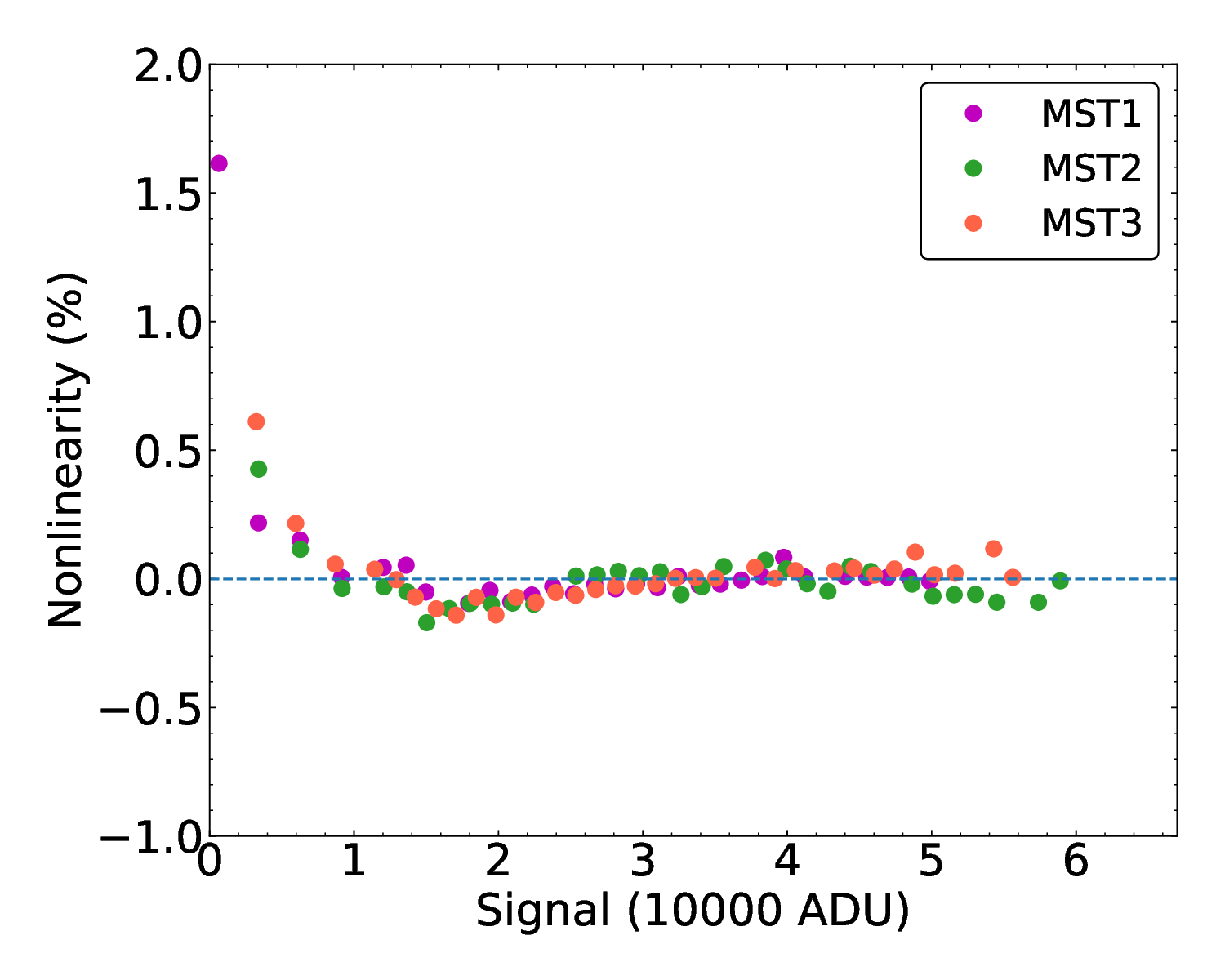}
\centering
\caption{Linearity assessment of of three ZWO ASI6200MM Pro CMOS cameras of three Mini-SiTian telescopes (MST1, MST2 and MST3). 
Left panel: Signal response (in units of 10000 ADU) versus exposure time (s) for 2000$\times$2000 pixels in the central region of the flat image. The linear trend is evident, with all cameras showing R-squared values exceeding 0.99997. 
Right panel: Nonlinearity percentage as a function of signal level. The plot reveals Minimal deviations from perfect linearity, particularly for signals above 3000 ADU, where nonlinearity remains consistently within $\pm$ 0.3\%. } \label{fig:linear} 
\end{figure}

Linearity is a critical parameter for evaluating the performance of a camera, as it determines how well the output signal corresponds to the input intensity. 
To examine the linearity of the ZWO ASI6200MM Pro cameras, a series of flat images were captured with exposure times ranging from 0.01 seconds to 8.1 seconds. 
Initially, it was found that the actual exposure time did not match the set exposure time, potentially affecting the assessment of linearity. Following communication with the manufacturer, ZWO, the firmware and drivers of the three cameras were updated to rectify this issue. 
Additionally, the manufacturer provided an Application Programming Interface to access the true exposure time, enabling our master control system (cite Wang) to record the accurate exposure time in the FITS file headers for the linearity assessment and further data processing. 
The left panel of Figure \ref{fig:linear} indicates the variation of the mean signal (in units of 10,000 ADU) for 2000$\times$2000 pixels in the central region of the flat image against exposure time. All three cameras exhibit strong linear trends, with R-squared values exceeding 0.99997, indicating excellent linearity across the tested range.
The right panel of Figure \ref{fig:linear} presents the nonlinearity percentage against Signal (in units of 10,000 ADU) for each camera, revealing the deviation from perfect linearity. 
The ZWO ASI6200MM Pro cameras exhibit different levels of nonlinearity across different signal ranges. For signals below 3000 ADU, although the Minimal signals may result in relatively large nonlinearities due to exposure time error, electronic noise, temperature fluctuations, or intrinsic sensor characteristics. Under most signal levels have nonlinearities below $\pm$ 0.5\%. 
The ZWO ASI6200MM Pro cameras exhibit exceptional performance characteristics for signals over 3000 ADU. The nonlinearity consistently remains within a remarkably low threshold of 0.3\%. This high degree of linearity is paramount for ensuring precise photometric measurements, a cornerstone of quantitative astronomical research. 
Such robust performance supports the suitability of the ZWO ASI6200MM Pro cameras for optical astronomical imaging applications, and the capacity to meet the exact standards required in precise astronomical observations.

\subsection{Gain and Read Noise} \label{sec:gain}
This section employs the conventional gain and read noise calculation method to determine the gain and read noise of ZWO ASI6200MM Pro cameras, while also introducing a novel approach for calculating pixel-level gain and read noise in CMOS sensors, which provides a more detailed characterization of CMOS sensor properties.

Gain and read noise are crucial parameters in evaluating the performance of CMOS cameras.
In CCD sensors, the gain is typically uniform across the entire sensor and can be calculated using a single value for the entire chip. 
This uniformity is due to the architecture of CCD, where charge is transferred through a common output amplifier.
The conventional method for calculating the gain of CCD is based on the photon transfer curve (PTC; \citealt{janesick2007dn}) technique, which uses the following formula:

\begin{equation} \label{eq:ccd_gain}
g = \frac{(\langle F_1\rangle - \langle B_1\rangle)+(\langle F_2\rangle - \langle B_2\rangle)}{\sigma_{F_{12}}^{2} - \sigma_{B_{12}}^{2}}
\end{equation}

where $\langle F_1\rangle$ and $\langle F_2\rangle$ represent the mean values of two flat-field images, respectively, $\langle B_1\rangle$ and $\langle B_2\rangle$ represent the mean values of two bias images, respectively, and $\sigma_{F_{12}}^{2}$ and $\sigma_{B_{12}}^{2}$ are the variances of the difference images (flat-flat and bias-bias, respectively).

In contrast, CMOS sensors have an amplifier for each pixel, resulting in pixel-to-pixel variations in gain. This fundamental difference requires a pixel-level approach to gain and readout noise calculation for CMOS sensors.

We propose a method to calculate the gain and read noise for each pixel. This method is based on the following assumptions:

\begin{enumerate}
    \item Flat-field images are obtained under stable light source conditions and with identical exposure times.
    \item Read noise and dark current noise remain stable over short periods.
\end{enumerate}

For a pixel at row $i$ and column $j$, the gain $g_{ij}$ is calculated using:

\begin{equation} \label{eq:gain}
g_{ij} = \frac{\langle F_{ij}\rangle - \langle B_{ij}\rangle}{\sigma_{F_{ij}}^2 - \sigma_{B_{ij}}^2}
\end{equation}

where $\langle F_{ij}\rangle$ and $\langle B_{ij}\rangle$ are the mean count values of the pixel in flat-field and bias images, respectively, and $\sigma_{F_{ij}}^2$ and $\sigma_{B_{ij}}^2$ are their corresponding variances.

The read noise $r_{ij}$ is then calculated as:

\begin{equation} \label{eq:read_noise}
r_{ij} = g_{ij}\sigma_{B_{ij}}
\end{equation}

The derivation of these formulas is based on the fundamental principles of the PTC. For an individual pixel, the total noise can be expressed as:

\begin{equation} \label{eq:total_noise}
(g_{ij}\sigma_{ij, \text{total}})^2 = r_{ij}^2 + g_{ij}\langle S_{ij}\rangle
\end{equation}

where $\langle S_{ij}\rangle$ represents the mean signal.

For bias images, where the signal is essentially zero, we have:

\begin{equation} \label{eq:bias_noise}
(g_{ij}\sigma_{B_{ij}})^2 = r_{ij}^2
\end{equation}

For flat-field images taken under stable light conditions and identical exposure times, we can write:

\begin{equation} \label{eq:flat_noise}
(g_{ij}\sigma_{F_{ij}})^2 = r_{ij}^2 + g_{ij}(\langle F_{ij}\rangle - \langle B_{ij}\rangle)
\end{equation}

Subtracting Equation (\ref{eq:bias_noise}) from Equation (\ref{eq:flat_noise}) yields:

\begin{equation}
g_{ij}^2(\sigma_{F_{ij}}^2 - \sigma_{B_{ij}}^2) = g_{ij}(\langle F_{ij}\rangle - \langle B_{ij}\rangle)
\end{equation}

which, when solved for $g_{ij}$, gives us Equation (\ref{eq:gain}).

We calculated the gain and read noise for each CMOS sensor using the conventional approach. The results are summarized in Table \ref{tab:gain_comparison}, which shows the gain and read noise values for the three MST cameras. The gain values are notably consistent across all cameras, ranging from 0.252 to 0.255 e$^-$ ADU$^{-1}$. This indicates similar sensitivity in converting incoming photons to digital signals. The read noise values show slight variations, with MST2 having the lowest at 1.464 e$^-$, while MST1 and MST3 have slightly higher values of 1.583 e$^-$ and 1.588 e$^-$ respectively. All cameras demonstrate excellent low-noise performance, with read noise values well below 1.6 e$^-$. 
These results demonstrate the high quality and consistency of the CMOS sensors used in the MST cameras.

To verify and compare our novel method, we applied it to the camera of MST3. We had access to 144 flat-field images taken with the same exposure time (2 seconds), which allowed us to demonstrate the application of our new method. 
Due to the limited flat images from MST1 and MST2, the resulting errors are large, so they will not be discussed in the paper.

\begin{figure}[htbp]
    \centering
    \includegraphics[width=\textwidth]{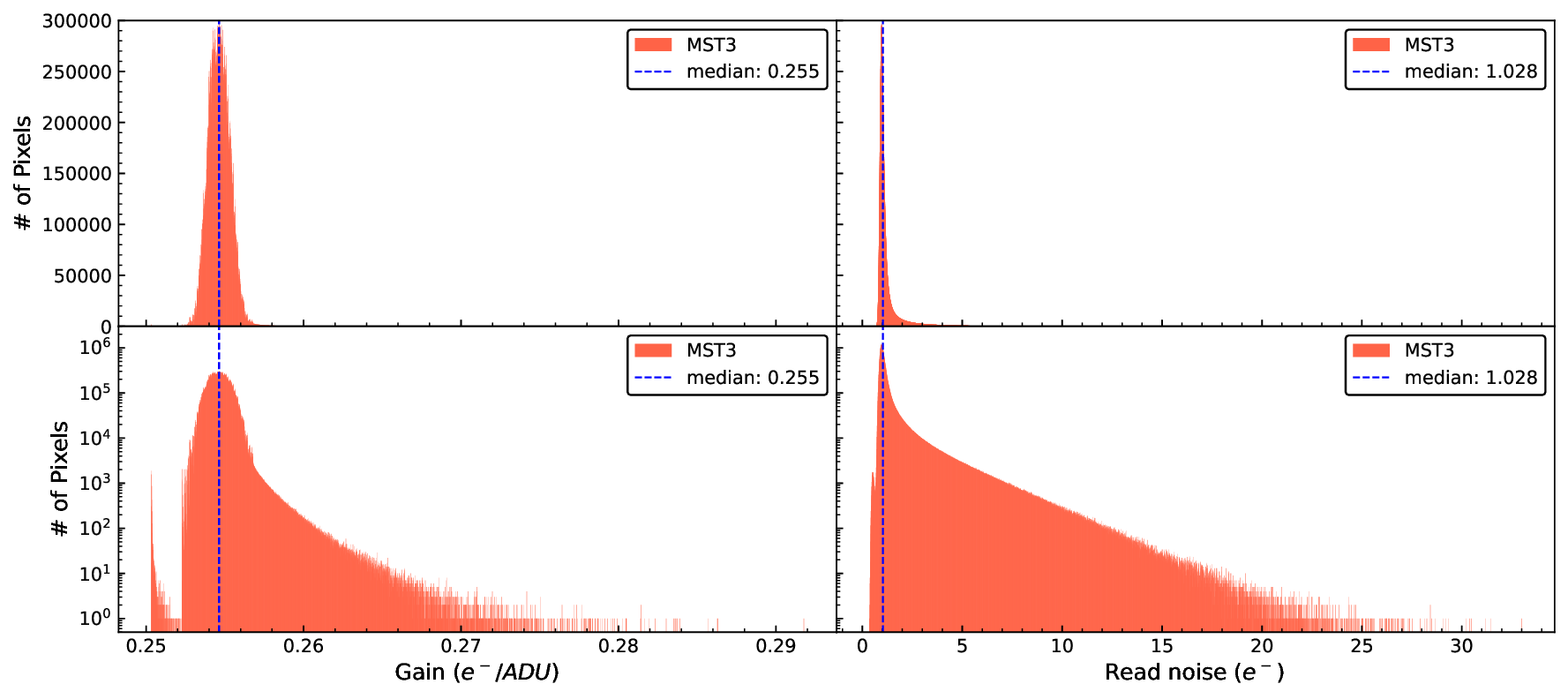}
    \caption{Pixel-level gain and read noise distribution for MST3. The top panels show the histograms in linear scale, while the bottom panels employ a logarithmic scale to highlight the distribution tails. The left panels illustrate the distribution of gain across pixels of the sensor, whereas the right panels depict the distribution of read noise across the pixels of the sensor. }
    \label{fig:mst3_gain_readnoise}
\end{figure}

Figure \ref{fig:mst3_gain_readnoise} presents a comprehensive analysis of the pixel-level gain and read noise characteristics of the MST3 CMOS sensor. 
The upper panels showing linear-scale histograms and the lower panels displaying the same data on a logarithmic scale to enhance visibility of low-frequency occurrences.

\begin{table}[!ht]
\centering
\caption{Gain and read noise values for the MST1, MST2, and MST3 cameras calculated using the conventional method} \label{tab:gain_comparison}
\begin{threeparttable}
\begin{tabular}{ccc}
\hline
Camera & Gain & Read Noise \\
 & (e$^-$ ADU$^{-1}$) & (e$^-$) \\
\hline
MST1 & 0.253 & 1.583 \\
MST2 & 0.252 & 1.464 \\
MST3 & 0.255 & 1.588 \\
\hline
\end{tabular}
\end{threeparttable}
\end{table}

The left panels illustrate the distribution of gain across pixels of the sensor, measured in electrons per ADU (e$^-$ ADU$^{-1}$). As shown in Figure \ref{fig:mst3_gain_readnoise}, the gain distribution for MST3 exhibits a bimodal characteristic. 
The primary peak is centered around a median of 0.255 e$^-$ ADU$^{-1}$, indicated by a blue dashed line, while a secondary, smaller peak is observed in the range of 0.250-0.252 e$^-$ ADU$^{-1}$.
The primary distribution is notably narrow and slightly right-skewed, with the majority of pixels clustered tightly around the median. This suggests a high degree of uniformity in the gain characteristics of the sensor, which is crucial for consistent image quality and accurate photometric measurements. 
The logarithmic plot in the lower left panel reveals a long tail extending towards higher gain values, indicating the presence of a small number of pixels with atypically high gain.
The presence of a bimodal distribution suggests two distinct populations of pixels with different gain characteristics. This bimodality is an important feature that may have implications for the performance of the sensor and calibration requirements.

The right panels depict the distribution of read noise across the pixels of the sensor, measured in electrons (e$^-$). 
The read noise distribution has a median of 1.028 e$^-$, shown by a blue dashed line. 
The distribution is characterized by two prominent peaks: a sharp, lower peak between 0 and 1 e$^-$, and a broader, higher peak centered around 1 e$^-$.
This bimodal structure suggests the presence of two distinct populations of pixels with slightly different noise characteristics.
The primary distribution is markedly asymmetric, with a sharp rise on the left side and a gradual decline on the right, forming a positively skewed distribution.
This low median read noise demonstrates the capability of the sensor for high-sensitivity imaging, particularly beneficial for faint object astronomical observations. 
The logarithmic plot in the lower right panel unveils a substantial tail extending towards higher read noise values, suggesting the presence of a subset of pixels with significantly elevated noise levels. 
This bimodal distribution is similar to the distribution of standard deviation value of each pixel derived from a series of bias images (see Figure \ref{fig:bad pix2}).

The gain results from the conventional method are generally consistent with the pixel-level analysis for MST3. However, it's important to note that the conventional method provides only a single value for the entire central region, while our pixel-level approach reveals the distribution of gain across the sensor. 

The difference in read noise can be attributed primarily to the inherent limitations of the conventional method and the unique characteristics of CMOS sensors. 
The conventional method provides an average value for the entire sensor, which is significantly influenced by Fixed Pattern Noise (FPN), a characteristic feature of CMOS sensors. 
FPN is a spatially fixed noise pattern that remains constant from frame to frame and is primarily caused by pixel-to-pixel variations in the sensor.

The conventional method, by calculating noise across the entire sensor, inherently includes the noise of FPN in its read noise estimation. This inclusion can lead to an overestimation of the read noise.
However, our pixel-level approach effectively isolates the impact of FPN by calculating gain and read noise for each pixel individually.

Overall, our pixel-level analysis method provides several advantages for characterizing CMOS cameras:

\begin{enumerate}
    \item Enhanced precision: By calculating gain and read noise for each individual pixel, we achieve a substantially higher level of precision compared to conventional methods. Traditional approaches typically provide only average values and inadvertently include the impact of FPN when calculating read noise. Our method circumvents these limitations, offering a more accurate representation of the camera's performance.
    \item Comprehensive uniformity assessment: The pixel-level distributions of gain and read noise generated by our method enable a detailed and nuanced assessment of sensor uniformity. This approach reveals the pixel response non-uniformity (PRNU; \citealt{jimenez2012high}) component of FPN and identifies anomalous pixels across the entire sensor array, providing invaluable insights into the sensor's spatial characteristics.
    \item Improved calibration capabilities: The pixel-level gain data obtained through our method can be effectively utilized to correct the PRNU component of FPN. Furthermore, the pixel-level read noise measurements allow for a more accurate noise assessment. 
    These refined measurements can significantly enhance the accuracy of photometric analyses, leading to more reliable and precise scientific results.
\end{enumerate}

The proposed method for calculating pixel-level gain and read noise provides a comprehensive characterization of CMOS camera performance. 
Our results demonstrate that the ZWO ASI6200MM Pro cameras exhibit good uniformity in gain across the sensor, with a narrow distribution centered around a specified value. The read noise distribution reveals a slightly skewed pattern, with a long tail towards higher values.
This detailed understanding of camera characteristics is essential for optimizing data reduction processes and achieving high-precision photometry in astronomical observations.

\section{Conclusion} \label{sec:conclusion}

This study evaluates the performance of the camera ZWO ASI6200MM Pro in a laboratory setting. Our findings demonstrate that the camera exhibits low dark currents, along with good short-term stability and excellent linearity, consistent with the specifications provided by the manufacturer. The key conclusions are as follows:

1. Bias Stability Analysis and Frame Characteristics: Our analysis of the ZWO ASI6200MM Pro CMOS cameras in the Mini-SiTian Array Camera System reveals several key characteristics. The cameras demonstrate good bias stability, with mean bias counts remaining within $\pm$0.03 ADU over a 20-minute period, and the deviation mean counts of bias in one night was less than 0.08 ADU (\citealt{2024RAA....24e5009M}). Combined bias frames exhibit subtle but consistent Fixed Pattern Noise (FPN), including gradients, corner effects, and horizontal structures. Notably, pixel value distributions in bias frames deviate from Gaussian, showing asymmetry with long tails towards higher ADU values. While all three cameras share these similar characteristics, slight variations exist between them. 

2. Anomalous response pixels: Our analysis of the ZWO ASI6200MM Pro CMOS cameras revealed a small but notable proportion of anomalous response pixels, ranging from 0.06\% to 0.08\%, with approximately 1.05\% showing significant variations in response. These pixels, predominantly located near the sensor edges and corners, exhibit characteristics similar to the performance of the anomalous response pixels of QHY411M \citep{alarcon_scientific_2023}. 
While the majority of pixels behave consistently and normally in both response and response deviation, the presence of these anomalous pixels underscores the importance of precise identification and annotation to maintain high photometric accuracy in astronomical observations.

3. Dark current: The ZWO ASI6200MM Pro CMOS cameras demonstrate good performance with a low average dark current rate of approximately 0.002 e$^-$ pixel$^{-1}$ s$^{-1}$ at $0^\circ\text{C}$. 
This low rate enhances the ability of the camera to detect faint celestial objects and improves image quality, particularly during long exposures. Our analysis revealed a non-uniform distribution of dark current across pixels. Future work will focus on characterizing this distribution in detail and developing advanced correction methods to further improve the overall system performance.

4. Linearity: Our analysis of the ZWO ASI6200MM Pro cameras reveals exceptional linearity across a wide range of exposure times and signal levels. The cameras consistently demonstrate nonlinearity below $\pm$ 0.5\%, with even higher performance (within 0.3\%) for signals above 3000ADU. This high degree of linearity, coupled with the consistent performance across multiple units, makes the ZWO ASI6200MM Pro an excellent choice for precision photometric measurements in various astronomical applications. 

5. Gain and read noise: We propose a novel method for calculating pixel-level gain and read noise in CMOS sensors, providing a more detailed characterization than conventional methods. Our analysis of the ZWO ASI6200MM Pro cameras reveals consistent gain values across all three MST cameras (0.252 to 0.255 e$^-$ ADU$^{-1}$). 
The pixel-level analysis for MST3 shows a narrow gain distribution centered around 0.255 e$^-$ ADU$^{-1}$, indicating excellent uniformity across the sensor. The read noise distribution exhibits a low median of 1.028 e$^-$. 
While most pixels perform consistently, the extended tails in both distributions reveal the presence of outlier pixels, which are important to identify for precise calibration. This detailed characterization enables more accurate calibration and optimized data reduction, crucial for high-precision photometric measurements in time-domain surveys. 

In conclusion, our comprehensive analysis of the ZWO ASI6200MM Pro CMOS camera demonstrates its potential for astronomical observations. Laboratory tests reveal characteristics such as low dark current, excellent linearity, good short-term stability, uniform gain, and low read noise, which suggest that the camera has the potential to meet the requirements of time-domain surveys.

The positive on-sky results from Xiao et al. (2024; also submitted to this special issue) further validate these findings. Their work confirms that the ZWO ASI6200MM Pro systems achieve high-precision photometric performance in real astronomical conditions, consistent with our laboratory observations. These results underscore the potential of these CMOS detectors for wide-field optical time-domain surveys, demonstrating photometric accuracy comparable to CCDs and impressive astrometric precision.

\normalem
\begin{acknowledgements}

This research is supported by National Key R\&D Program of China (Grant Nos. 2023YFA1609700, 2023YFA1608304, and 2023YFA1608303), the Strategic Priority Research Program of the Chinese Academy of Sciences (XDB0550103), and the National Natural Science Foundation of China (NSFC, Nos. 12090040 and 12090041).

The SiTian project is a next-generation, large-scale time-domain survey designed to build an array of over 60 optical telescopes, primarily located at observatory sites in China. This array will enable single-exposure observations of the entire northern hemisphere night sky with a cadence of only 30-minute, capturing true color (gri) time-series data down to about 21 mag. This project is proposed and led by the National Astronomical Observatories, Chinese Academy of Sciences (NAOC). As the pathfinder for the SiTian project, the Mini-SiTian project utilizes an array of three 30 cm telescopes to simulate a single node of the full SiTian array. The Mini-SiTian has begun its survey since Novmber 2022. The SiTian and Mini-SiTian have been supported from the Strategic Pioneer Program of the Astronomy Large-Scale Scientific Facility, Chinese Academy of Sciences and the Science and Education Integration Funding of University of Chinese Academy of Sciences.

We are grateful to Dr. Xianmin Meng, Liguo Fang, Jiupeng Guo, Wan Zhou, and Qiuyan Luo for their helpful assistance with the experimental setup.

\end{acknowledgements}

\vspace{5mm}
\facilities{Mini-SiTian telescope (XingLong Observatory, NAOC)}

\software{Matplotlib \citep{Hunter:2007}, Numpy \citep{harris2020array}, Astropy \citep{2013A&A...558A..33A, 2018AJ....156..123A}
}

\bibliographystyle{raa}
\bibliography{mst}

\begin{thebibliography}{20}
\providecommand\natexlab[1]{#1}
\providecommand\JournalTitle[1]{#1}

\bibitem[Alarcon {et~al.}(2023)]{alarcon_scientific_2023}
Alarcon, M.~R., Licandro, J., Serra-Ricart, M., {et~al.} 2023, Publications of
  the Astronomical Society of the Pacific, 135, 055001, aDS Bibcode:
  2023PASP..135e5001A

\bibitem[{Astropy Collaboration} {et~al.}(2013)]{2013A&A...558A..33A}
{Astropy Collaboration}, {Robitaille}, T.~P., {Tollerud}, E.~J., {et~al.} 2013,
  \aap, 558, A33

\bibitem[{Astropy Collaboration} {et~al.}(2018)]{2018AJ....156..123A}
{Astropy Collaboration}, {Price-Whelan}, A.~M., {Sip{\H{o}}cz}, B.~M., {et~al.}
  2018, \aj, 156, 123

\bibitem[Bigas {et~al.}(2006)]{bigas_review_2006}
Bigas, M., Cabruja, E., Forest, J., \& Salvi, J. 2006, Microelectronics
  Journal, 37, 433

\bibitem[{Duan} {et~al.}(2020)]{2020AcASn..61...37D}
{Duan}, W., {Song}, Q., {Bai}, X.~Y., {et~al.} 2020, Acta Astronomica Sinica,
  61, 37

\bibitem[Gamal {et~al.}(1998)]{ElGamal1998ModelingAE}
Gamal, A.~E., Fowler, B., Min, H., \& Liu, X. 1998, in Electronic imaging

\bibitem[{Gunn} {et~al.}(1998)]{1998AJ....116.3040G}
{Gunn}, J.~E., {Carr}, M., {Rockosi}, C., {et~al.} 1998, \aj, 116, 3040

\bibitem[Harris {et~al.}(2020)]{harris2020array}
Harris, C.~R., Millman, K.~J., van~der Walt, S.~J., {et~al.} 2020, Nature, 585,
  357

\bibitem[Hunter(2007)]{Hunter:2007}
Hunter, J.~D. 2007, Computing in Science \& Engineering, 9, 90

\bibitem[Janesick(2007)]{janesick2007dn}
Janesick, J. 2007, DN to [lambda]:, Press Monographs (SPIE)

\bibitem[Jim{\'e}nez-Garrido {et~al.}(2012)]{jimenez2012high}
Jim{\'e}nez-Garrido, F., Fern{\'a}ndez-P{\'e}rez, J., Utrera, C., {et~al.}
  2012, in Sensors, Cameras, and Systems for Industrial and Scientific
  Applications XIII, ed. R.~Widenhorn, V.~Nguyen, \& A.~Dupret, Vol. 8298,
  International Society for Optics and Photonics (SPIE), 829803

\bibitem[{Karpov} {et~al.}(2021)]{2021RMxAC..53..190K}
{Karpov}, S., {Christov}, A., {Bajat}, A., {Cunniffe}, R., \& {Prouza}, M.
  2021, in Revista Mexicana de Astronomia y Astrofisica Conference Series,
  Vol.~53, Revista Mexicana de Astronomia y Astrofisica Conference Series, 190

\bibitem[Konnik \& Welsh(2014)]{Konnik2014HighlevelNS}
Konnik, M.~V., \& Welsh, J.~S. 2014, ArXiv, abs/1412.4031

\bibitem[{Liu} {et~al.}(2021)]{2021AnABC..93..628L}
{Liu}, J., {Soria}, R., {Wu}, X.-F., {Wu}, H., \& {Shang}, Z. 2021, Anais da
  Academia Brasileira de Ciencias, 93, 20200628

\bibitem[{Ma} {et~al.}(2012)]{2012SPIE.8446E..6RM}
{Ma}, B., {Shang}, Z., {Wang}, L., {et~al.} 2012, in Society of Photo-Optical
  Instrumentation Engineers (SPIE) Conference Series, Vol. 8446, Ground-based
  and Airborne Instrumentation for Astronomy IV, ed. I.~S. {McLean}, S.~K.
  {Ramsay}, \& H.~{Takami}, 84466R

\bibitem[{Mu} {et~al.}(2024)]{2024RAA....24e5009M}
{Mu}, H.-Y., {Fan}, Z., {Zhu}, Y.-N., {Zhang}, Y., \& {Wu}, H. 2024, Research
  in Astronomy and Astrophysics, 24, 055009

\bibitem[{Princeton Instruments}(2016)]{PCO_scmos_white_paper}
{Princeton Instruments}. 2016, New Scientific CMOS Cameras with
  Back-Illuminated Technology,
  \url{https://www.princetoninstruments.com/wp-content/uploads/2020/11/TechNote_sCMOSBackIlluminatedTech.pdf},
  accessed: July 30, 2024

\bibitem[Qiu {et~al.}(2013)]{qiu_evaluation_2013}
Qiu, P., Mao, Y.-N., Lu, X.-M., Xiang, E., \& Jiang, X.-J. 2013, Research in
  Astronomy and Astrophysics, 13, 615

\bibitem[Wang {et~al.}(2017)]{Wang_2017}
Wang, L., Ma, B., Li, G., {et~al.} 2017, The Astronomical Journal, 153, 104

\bibitem[Zhang {et~al.}(2020)]{zhang_tsinghua_2020}
Zhang, J.-C., Wang, X.-F., Mo, J., {et~al.} 2020, Publications of the
  Astronomical Society of the Pacific, 132, 125001

\end{thebibliography}

\end{document}